\documentclass[12pt,preprint]{aastex}

\usepackage{lscape}
\usepackage{amsmath}

\usepackage{epstopdf}
\usepackage{color}

\slugcomment{Accepted by ApJS}

\shorttitle{BLAST observations of the SEP field}
\shortauthors{Valiante et al.}

\begin{document}


\title{BLAST observations of the South Ecliptic Pole field: number counts and source catalogs}
\author{
        Elisabetta~Valiante \altaffilmark{1,\dag},
        Peter~A.~R.~Ade \altaffilmark{2},
        James~J.~Bock \altaffilmark{3,4},
        Filiberto~G.~Braglia \altaffilmark{1},
        Edward~L.~Chapin \altaffilmark{1},
        Mark~J.~Devlin \altaffilmark{5},
        Matthew~Griffin \altaffilmark{2},
        Joshua~O.~Gundersen \altaffilmark{6},
        Mark~Halpern \altaffilmark{1},
        Peter~C.~Hargrave \altaffilmark{2},
        David~H.~Hughes \altaffilmark{7},
        Jeff~Klein \altaffilmark{5},
        Gaelen~Marsden \altaffilmark{1},
        Philip~Mauskopf \altaffilmark{2},
        Calvin~B.~Netterfield \altaffilmark{8,9},
        Luca~Olmi \altaffilmark{10,11},
        Enzo~Pascale \altaffilmark{2},
        Guillaume~Patanchon \altaffilmark{12},
	Marie~Rex \altaffilmark{13},
        Douglas~Scott \altaffilmark{1},
        Kimberly~Scott \altaffilmark{5},
        Christopher~Semisch \altaffilmark{5},
	Hans~Stabenau \altaffilmark{5},
        Nicholas~Thomas \altaffilmark{6},
        Matthew~D.~P.~Truch \altaffilmark{5},
        Carole~Tucker \altaffilmark{2},
        Gregory~S.~Tucker \altaffilmark{14},
        Marco~P.~Viero \altaffilmark{9},
        Donald~V.~Wiebe \altaffilmark{1}
}

\altaffiltext{1}{Department of Physics \& Astronomy, University of British Columbia, 6224 Agricultural Road, Vancouver, BC V6T~1Z1, Canada}
\altaffiltext{2}{Department of Physics \& Astronomy, Cardiff University, 5 The Parade, Cardiff, CF24~3AA, UK}
\altaffiltext{3}{Jet Propulsion Laboratory, Pasadena, CA 91109-8099, USA}
\altaffiltext{4}{Observational Cosmology, MS 59-33, California Institute of Technology, Pasadena, CA 91125, USA}
\altaffiltext{5}{Department of Physics \& Astronomy, University of Pennsylvania, 209 South 33rd Street, Philadelphia, PA 19104, USA}
\altaffiltext{6}{Department of Physics, University of Miami, 1320 Campo Sano Drive, Carol Gables, FL 33146, USA}
\altaffiltext{7}{Instituto Nacional de Astrof\'isica \'Optica y Electr\'onica (INAOE), Aptdo. Postal 51 y 72000 Puebla, Mexico}
\altaffiltext{8}{Department of Astronomy \& Astrophysics, University of Toronto, 50 St. George Street, Toronto, ON  M5S~3H4, Canada}
\altaffiltext{9}{Department of Physics, University of Toronto, 60 St. George Street, Toronto, ON M5S~1A7, Canada}
\altaffiltext{10}{Istituto di Radioastronomia, Largo E. Fermi 5, I-50125, Firenze, Italy}
\altaffiltext{11}{University of Puerto Rico, Rio Piedras Campus, Physics Dept., Box 23343, UPR station, San Juan, Puerto Rico}
\altaffiltext{12}{Laboratoire APC, 10, rue Alice Domon et L{\'e}onie Duquet 75205 Paris, France}
\altaffiltext{13}{Steward Observatory, University of Arizona, 933 N. Cherry Ave, Tucson, AZ 85721, USA}
\altaffiltext{14}{Department of Physics, Brown University, 182 Hope Street, Providence, RI 02912, USA}
\altaffiltext{\dag}{\url{valiante@phas.ubc.ca}}

\begin{abstract}
We present results from a survey carried out by the Balloon-borne Large Aperture Submillimeter Telescope (BLAST) on a $9\,{\rm deg}^2$ field near the South Ecliptic Pole at 250, 350 and $500\,\micron$. The median $1\sigma$ depths of the maps are 36.0, 26.4 and $18.4\,{\rm mJy}$, respectively. We apply a statistical method to estimate submillimeter galaxy number counts and find that they are in agreement with other measurements made with the same instrument and with the more recent results from {\itshape Herschel}/SPIRE. Thanks to the large field observed, the new measurements give additional constraints on the bright end of the counts. We identify 132, 89 and 61 sources with S/N $\geq 4$ at 250, 350, $500\,\micron$, respectively and provide a multi-wavelength combined catalog of 232 sources with a significance $\geq 4\sigma$ in at least one BLAST band. The new BLAST maps and catalogs are available publicly at http://blastexperiment.info.  
\end{abstract}


\section{Introduction}
Understanding the formation and evolution of galaxies is one of the foremost goals of experimental cosmology. In the redshift range $z\simeq 1-3$, massive galaxies go through an evolutionary stage characterized by high rates of star formation. These early, dusty galaxies are best characterized by their thermal dust emission at far-infrared and submillimeter wavelengths. They are known to be the main component of the Cosmic Infrared Background (CIB; \citealt{fixsen98,hauser01}), but still relatively little is known about their nature and evolution.

Observations by the Balloon-borne Large Aperture Submillimeter Telescope (BLAST; \citealt{pascale08}) have provided the first confusion-limited submillimeter maps at 250, 350 and $500\,\micron$, with a beam size of $36^{\prime\prime}$, $42^{\prime\prime}$ and $60^{\prime\prime}$, respectively, covering areas larger than $1\,{\rm deg}^2$. BLAST carried out surveys of the extragalactic sky in two blank fields, one centered on the southern field of the Great Observatories Origins Deep Survey (GOODS-South; \citealt{devlin09}) and one close to the South Ecliptic Pole (SEP). Each survey covers about $10\,{\rm deg}^2$. 

BLAST is the first instrument to provide maps of the sky at wavelengths near the peak of the CIB with enough sensitivity, sky coverage and angular resolution to identify a large number of sources, determine the detailed shape of the source counts and show that most of the FIRB comes from submillimeter sources already identified in deep $24\,\micron$ surveys \citep{devlin09,patanchon09,marsden09}. There have been several studies \citep{marsden09,pascale09} of the statistical properties in the BLAST bands of sources from catalogs defined using {\itshape Spitzer} $24\,\micron$ observations (FIDEL, \citealt{dickinson07}). These have shown that $24\,\micron$ sources may well contribute all of the CIB, with about half of the emission coming from galaxies at $z\gtrsim 1.2$. Other statistical analyses of the BLAST GOODS-South (BGS) data \citep{patanchon09,viero09} have provided measurements of the differential number counts and of clustering on scales larger than the BLAST beam.  

In this paper we present results from the BLAST survey close to the SEP (hereafter BSEP field). This field was chosen because of very low emission from infrared cirrus, inferred from the IRAS $100\,\micron$ map of \citet{schlegel98}. Although point source detection (i.e. at high angular frequency) is not affected by fluctuations at large angular scales (of the order of a degree) from cirrus noise, a low-cirrus region is needed for studies of the CIB at these wavelengths. The SEP field does not have the same richness of multi-wavelength data as GOODS-South, but the multi-wavelength coverage is improving, in particular because of the observations carried out toward that region by the {\itshape AKARI} \citep{matsuhara06,malek10,matsuura10}, AzTEC on the Atacama Submillimeter Telescope Experiment (Hatsukade et al. in prep.), the South Pole Telescope, the Atacama Cosmology Telescope and the Australia Telescope Compact Array. The catalog presented in this paper can provide submillimeter spectral energy distributions (SEDs) for the AzTEC and {\itshape AKARI} sources. Moreover, the BSEP field is one of the HerMES\footnote{http://hermes.sussex.ac.uk} fields (ADF-S), so that the BSEP catalog can be used for cross-checking of the HerMES ADF-S data.

We have performed an analysis of the statistical properties of the maps to estimate number counts and we have extracted sources using the Point Spread Function (PSF) to filter the maps, applying the same techniques used in previous BLAST papers \citep{devlin09,patanchon09}. A more recent study by \cite{chapin10} proposes a different approach for filtering the submillimeter maps, namely a `matched filter' optimized to maximize the signal-to-noise ratio (S/N) of individual point sources in the presence of multiple noise sources. This approach significantly reduces the confusion noise, at the expense of slightly higher instrumental white noise. We have also used this approach for the BSEP data and have compared the results using both the PSF and the `matched filter'.

The same field has been also observed by {\itshape Spitzer}-MIPS at 24 and $70\,\micron$ \citep{scott10}. The combination of {\itshape Spitzer} and BLAST data enables the identification of mid-IR counterparts for $\sim 50\%$ of the BLAST sources and also provides useful limits for unidentified sources. The combined results can be used to constrain the SEDs of the galaxies across the entire Rayleigh-Jeans to Wien region of the far-IR spectrum and will be presented in a future paper.

The paper is organized as follows: \S~2 introduces the BLAST data; \S~3 presents the statistical analysis used to calculate differential number counts and a comparison between number counts calculated in this field with previous measurements, as well as comparisons with the most recent evolutionary models; \S~4 provides source lists constructed independently for each BLAST band and combined together and \S~5 gives some brief conclusions. The source catalogs are provided in the Appendix. This paper constitutes the public release of the BLAST maps and multiwavelength catalogs of the BSEP field.    

\section{BLAST Data}
\label{sec:BLASTdata}

\begin{figure}
\centering
\includegraphics[width=15.cm,keepaspectratio]{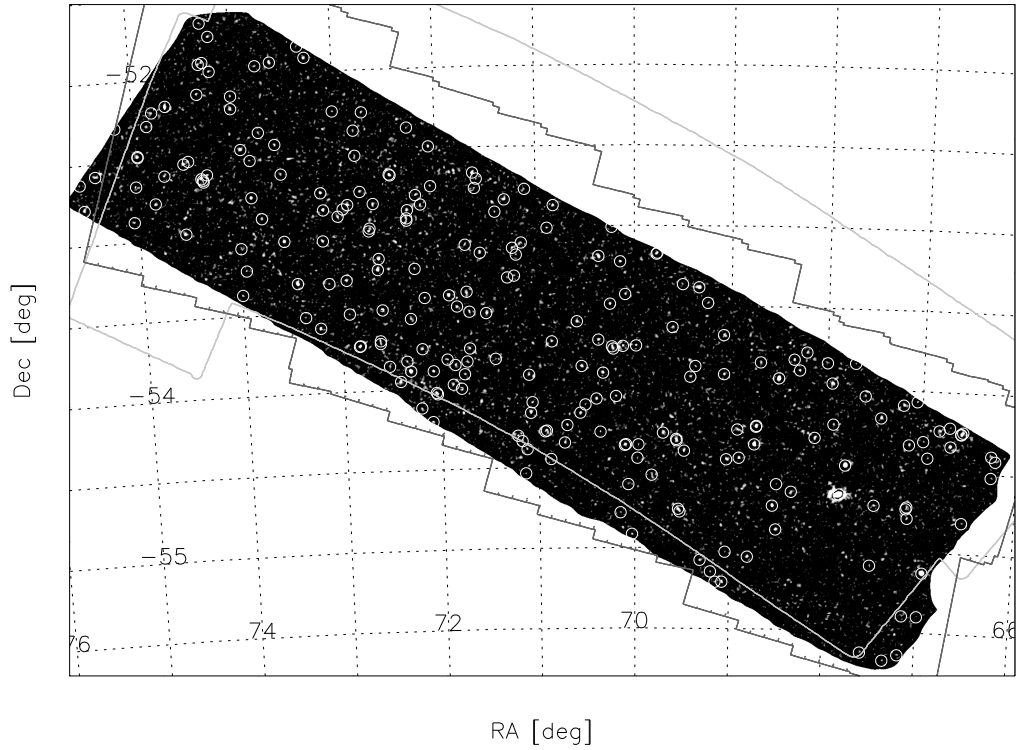}  
\caption{Map showing the combined signal-to-noise ratio of all three BLAST bands in the BSEP field. The circles mark the locations of the 232 sources detected with S/N $\geq 4$ in at least one band, as listed in Table~\ref{tab:BLASTcat}. The contours show the overlapping coverage of the $11.5\,{\rm deg}^2$ {\itshape Spitzer} survey in this field ({\itshape dark gray}) and the region mapped at $90\,\micron$ by {\itshape AKARI} ({\itshape light gray}).}
\label{fig:map}
\end{figure}

During the BLAST 2006 Antarctic flight, deep observations (68 hours) were carried out over a $9\,{\rm deg}^2$ field centered on ($70.94^{\circ}$,$-53.50^{\circ}$) near the SEP. The BSEP maps are made from a large number of cross-linked scans, producing a uniform map with median 1$\sigma$ sensitivities (equivalent for detecting point sources) of $36.0\,\rm{mJy}$, $26.4\,\rm{mJy}$, and $18.4\,\rm{mJy}$, for the 250, 350, and 500$\,\micron$ bands, respectively. Fig.~\ref{fig:map} shows the combined signal-to-noise ratio image of all three BLAST bands. These new maps cover an area comparable to the BGS-Wide map of \cite{devlin09}. We refer the reader to \citet{pascale08} for a more detailed description of the characteristics of BLAST and \citet{truch09} for information on calibration and data reduction.

The BLAST time-stream data were reduced using a common pipeline to identify spikes, correct detector time drift and calibrate data \citep{pascale08,truch09}. Maps were generated using the SANEPIC software, which uses a maximum-likelihood algorithm to estimate the optimal solution for the map, as well as producing an associated noise map \citep{patanchon08}. Absolute calibration is based on observations of the evolved star VY CMa and is estimated to have an uncertainty of 10\% (although strongly correlated between the three bands; see \citealt{truch09} for details).

While the maps represent the optimal weighting of the data across all spatial scales, the largest scales are less constrained due to various systematic effects. This can produce residual large-scale fluctuating patterns across the map. To suppress these spurious signals, all maps have been filtered to remove large scale angular frequencies, without affecting the contribution of individual point sources. This corresponds to scales in excess of about 10 arcmin (approximately the size of the detector array projected on the sky). This procedure, already used for the BGS maps (e.g.~\citealt{devlin09}), also explicitly sets the mean of each map to zero. The filtering induces negative shadows around the locations of bright sources in the maps, but this effect is fully taken into account in the analyses presented in this paper.

The confusion noise in the map is estimated with the method explained in \cite{marsden09}. We fit a Gaussian to the distribution of the ratio of map pixel values to the instrumental noise $\sigma_{\rm i}$. In the case of no confusion noise, the standard deviation of this Gaussian is unity by definition; if confusion noise is present, it gives the total noise of the map, $\sigma_{\rm t}$, in units of $\sigma_{\rm i}$. The confusion noise, $\sigma_{\rm c}$, is then derived by subtracting the instrumental noise from the total noise in quadrature, $\sigma_{\rm c}=\sqrt{\sigma_{\rm t}^2-\sigma_{\rm i}^2}$. Results are reported in Table~\ref{tab:catalogs}. As for the BGS-Wide map, the BSEP map is dominated by instrumental noise and so, using this criterion, the maps are not confusion limited.

The BLAST maps in this release include three primary data products. The first set contains the `raw' maps produced by SANEPIC. The second set of maps have been filtered spatially using a Wiener filter to whiten structures on scales larger than the instantaneous field of view. We also provide `matched-filtered' maps (see \S~\ref{sec:matched_filter}): these maps have been filtered using a `matched filter', which gives superior performs as maps approach the confusion limit \citep{chapin10}. 

\section{Number Counts}
\label{sec:BLASTnumbercounts}

\subsection{{\itshape P(D)} analysis}

Number counts are estimated using the same method adopted in \cite{patanchon09}. In that paper it was shown that, in the S/N regime of BLAST, a statistical analysis of the maps is a better approach for obtaining number counts than counting individual sources. This is because it naturally allows for the correction of strong biases due to confusion and flux boosting. This technique also has the advantage of providing an unbiased estimate of the counts at flux densities well below the limit at which sources can be detected individually. The method has been optimized to deal with inhomogeneous noise across the map and filtering to suppress large-scale noise.

We start with a measure of the `probability of deflection', {\itshape P(D)}, the histogram of pixel values, and use this to try to obtain the best estimate of the number counts. As in \cite{patanchon09}, we choose to parametrize the differential number counts by a set of amplitudes at a few predefined fluxes, with the intervals between flux nodes interpolated with power-laws to impose continuity of the counts. The number and locations of the nodes are chosen as a compromise between increasing the number of free parameters, to give a better representation of the true underlying counts, and keeping the number of nodes small enough to provide useful estimates of their values. We find that no more than about four amplitude parameters can be estimated for each waveband. This is lower than the numbers estimated from the BGS maps (up to seven nodes at $250\,\micron$), but this is not unexpected given the extra dynamic range of BGS-Wide plus BGS-Deep. Our observations provide useful information starting from $\sim 0.05\,{\rm Jy}$ and brighter. A larger number of nodes would increase the correlation between neighbouring nodes and thus would not add any information.

The {\itshape P(D)} analysis is carried out by minimizing the negative log-likelihood
\begin{equation} \label{eq:neglik}
\Phi(\theta)=-\sum_{i} n_i\log(p_i(\theta))-\log(N!)+\sum_{i} \log(n_i!) ,
\end{equation}  
where $n_i$ is the number of pixels with flux densities in the $i$th flux bin interval; $p_i$ is the result of the integral of the {\itshape P(D)} in the $i$th bin, normalized such that $\sum p_i=1$; $N$ is the total number of measurements (pixels in the map); $\theta$ are the parameters of the number counts model and the last two terms derive from the normalization of the multinomial distribution function (see \citealt{patanchon09} for a derivation). Eq.~\ref{eq:neglik} is only strictly correct if the probability distribution is the same for all observations (pixels). This is not the case for the BGS maps, with the expected distributions being different for the deep and the wide regions, and it is not completely true for the BSEP maps, even though differences in the noise levels are not as dramatic as in the BGS case. To deal with this we have divided the observed maps into a small number of regions (4), such that in each zone the noise variance is approximately constant. We then compute the quantities in Eq.~\ref{eq:neglik} for each of the regions.

The {\itshape P(D)} analysis works under the hypothesis that all the sources are point sources, i.e. all sources in the map have the size of the beam. This is not exactly true in our BSEP maps, where at least one local galaxy is partly resolved by the instrument and appears as an extended source (see \S~\ref{sec:extended}). Before performing the fit, we have masked the extended source and its dark ring.

\begin{figure}

\centering
{\includegraphics[width=8.cm,keepaspectratio]{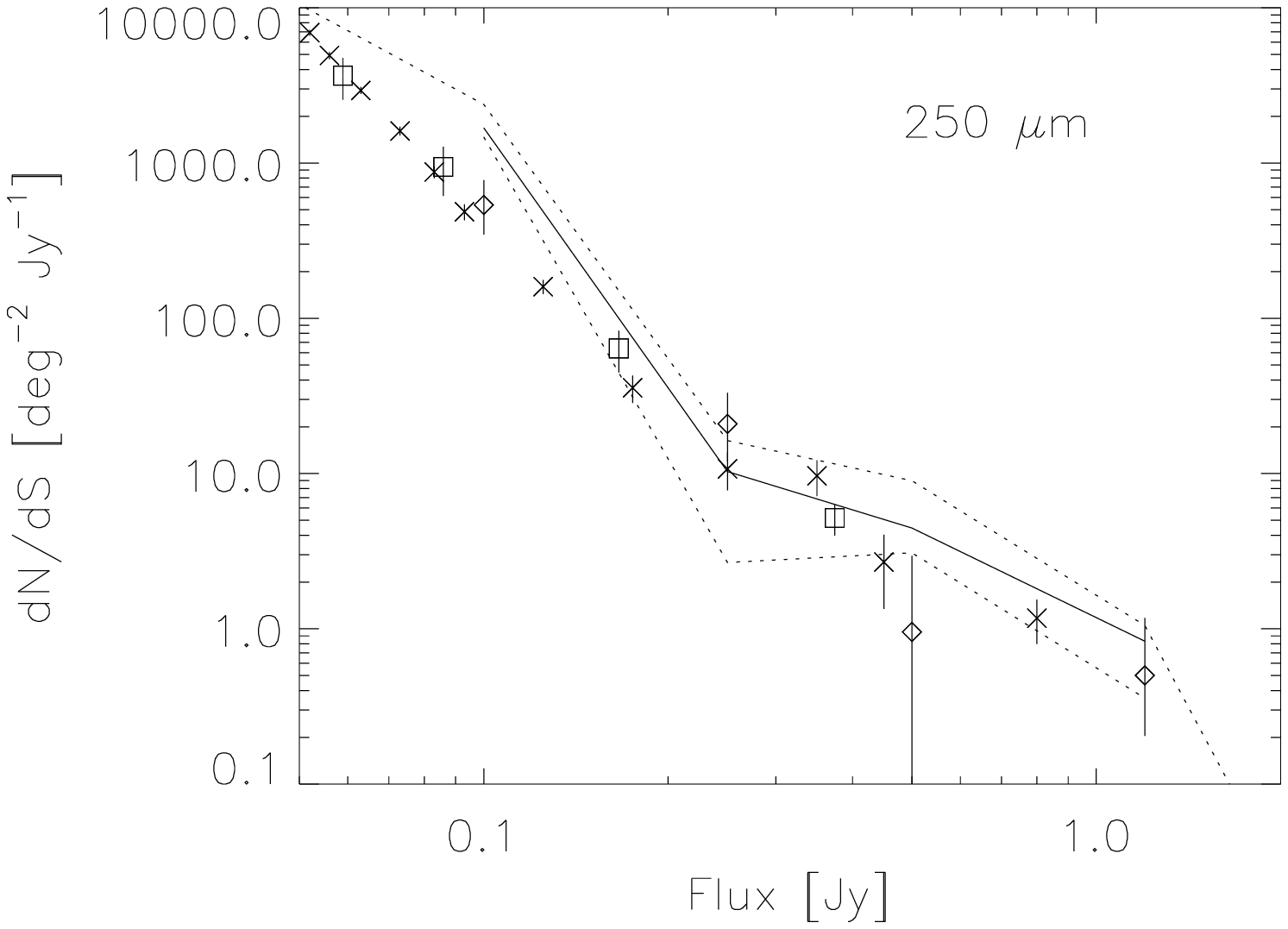}}
{\includegraphics[width=8.cm,keepaspectratio]{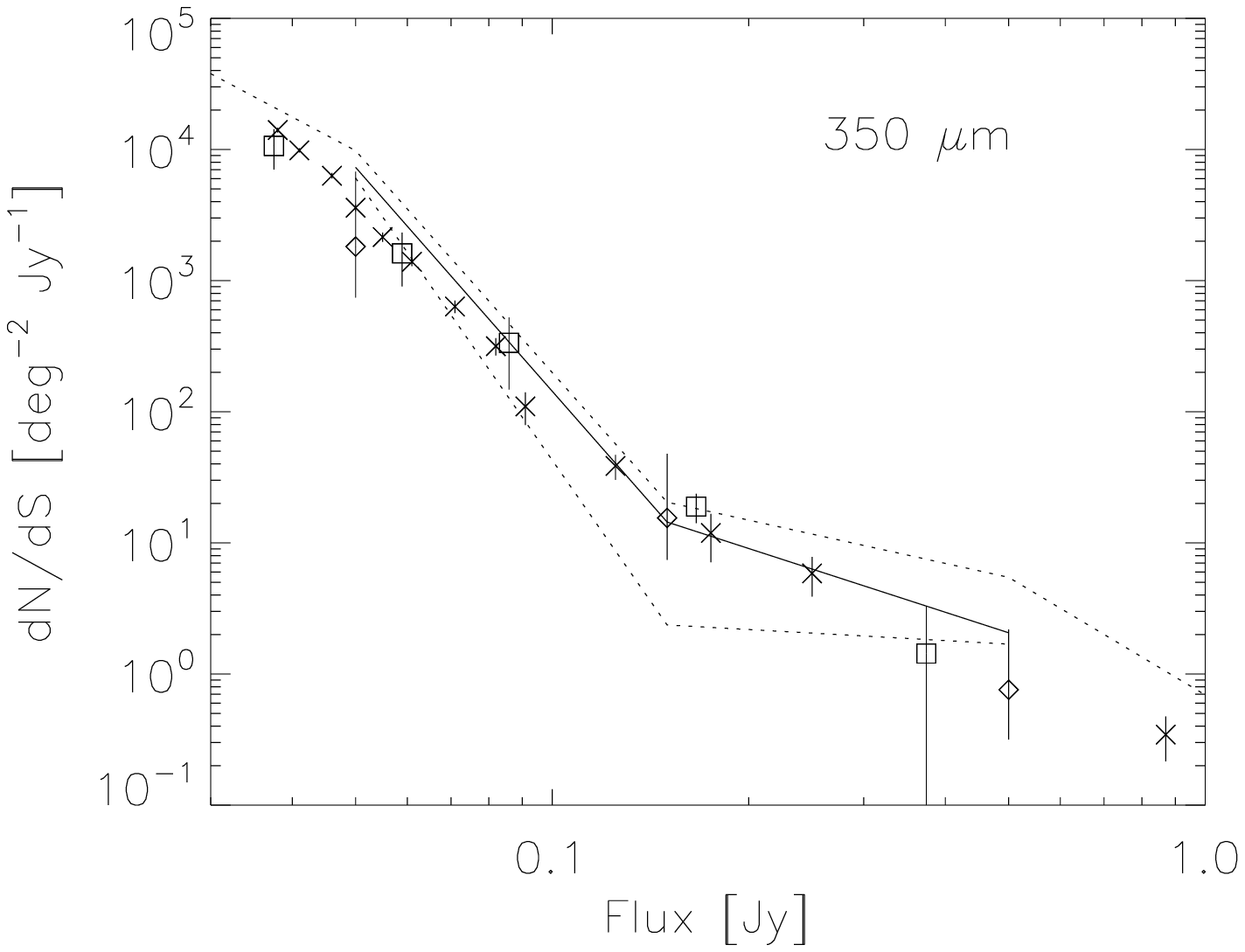}}\\
{\includegraphics[width=8.cm,keepaspectratio]{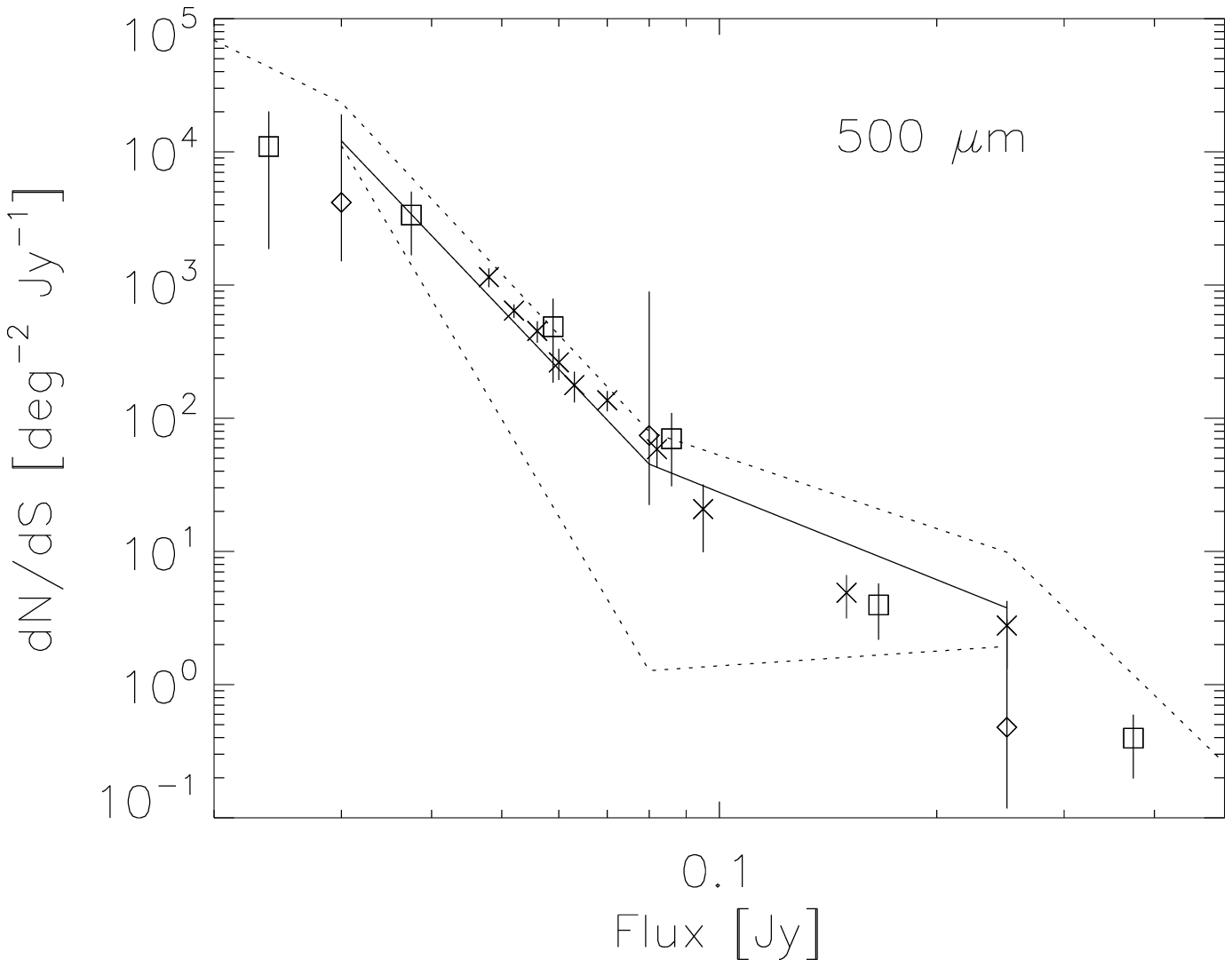}}
\caption{Best-fit differential number counts at the three BLAST wavelengths ({\itshape solid lines}). Uncertainties are derived from the marginalized distribution for each parameter ({\itshape dotted lines}); because of the non-Gaussian behavior of the likelihood around its maximum, the best-fit model is not centered on the error bars. The first and last power-laws ({\itshape dotted lines}) are upper limits. Number counts in the same bands from other studies are also plotted: \citet{patanchon09} ({\itshape diamonds}); \citet{oliver10} ({\itshape squares}); \citet{clements10} ({\itshape crosses}).
}
\label{fig:best_fit}
\end{figure}

\begin{figure}

\centering
{\includegraphics[width=8.cm,keepaspectratio]{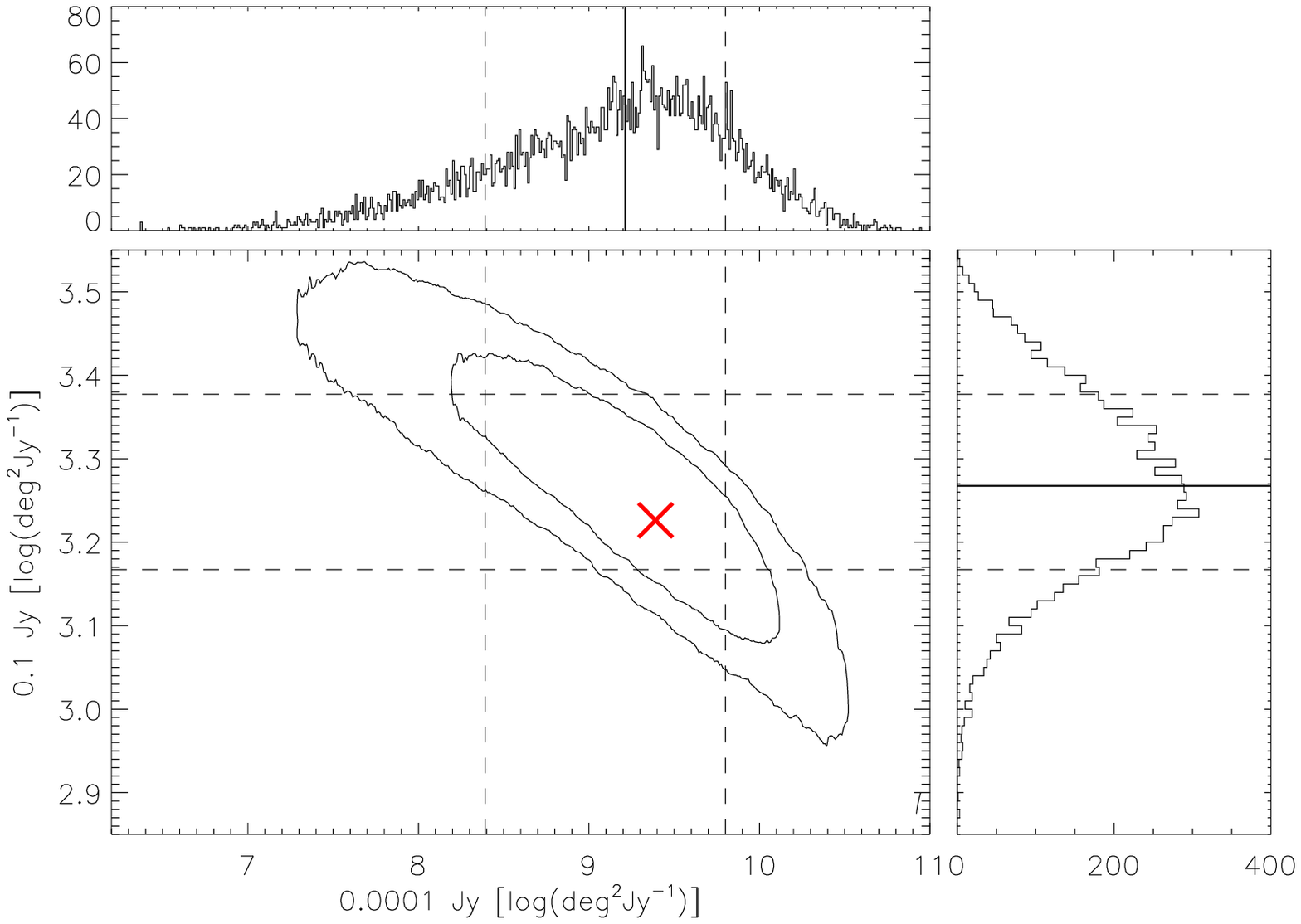}}
{\includegraphics[width=8.cm,keepaspectratio]{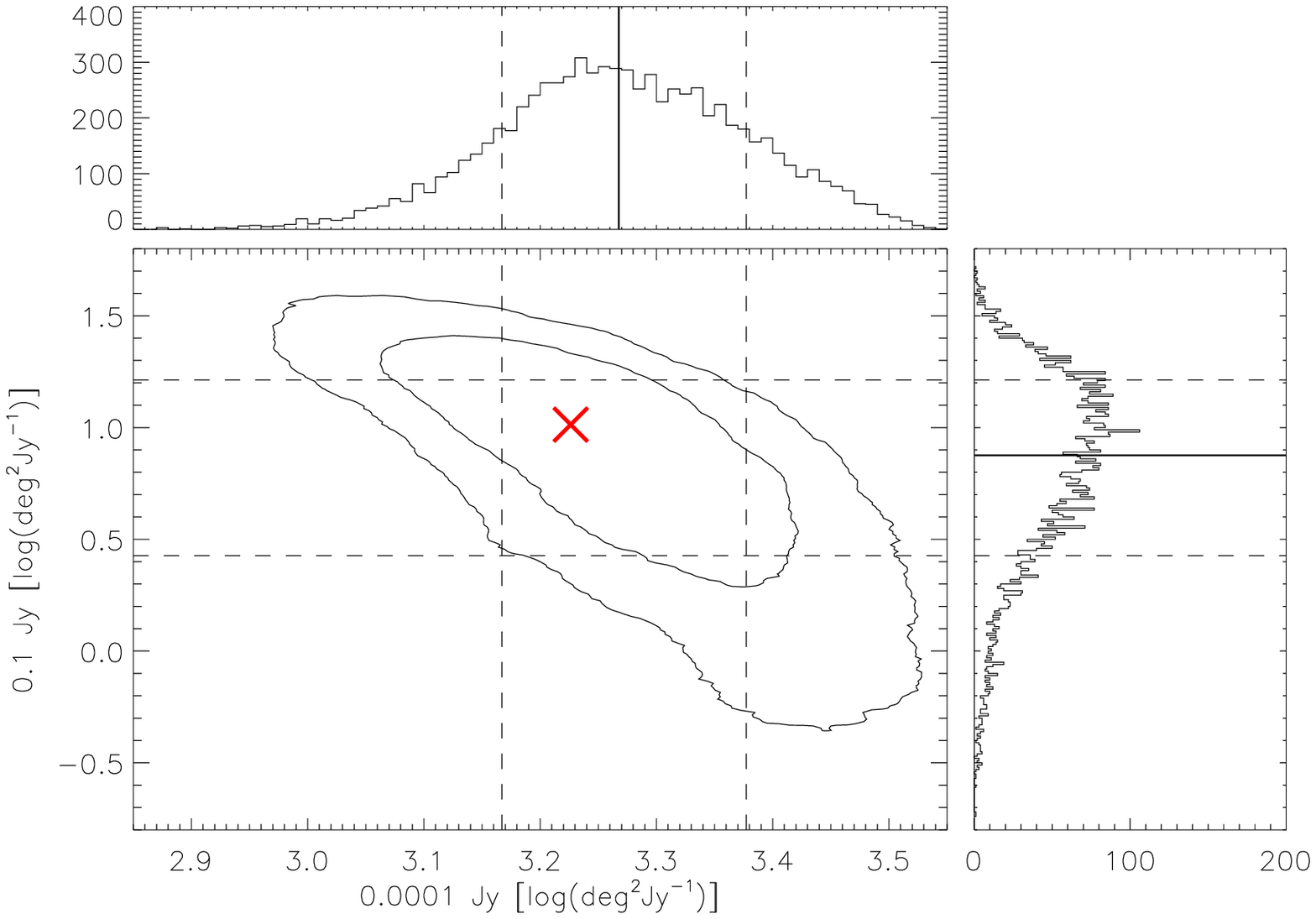}}\\
{\includegraphics[width=8.cm,keepaspectratio]{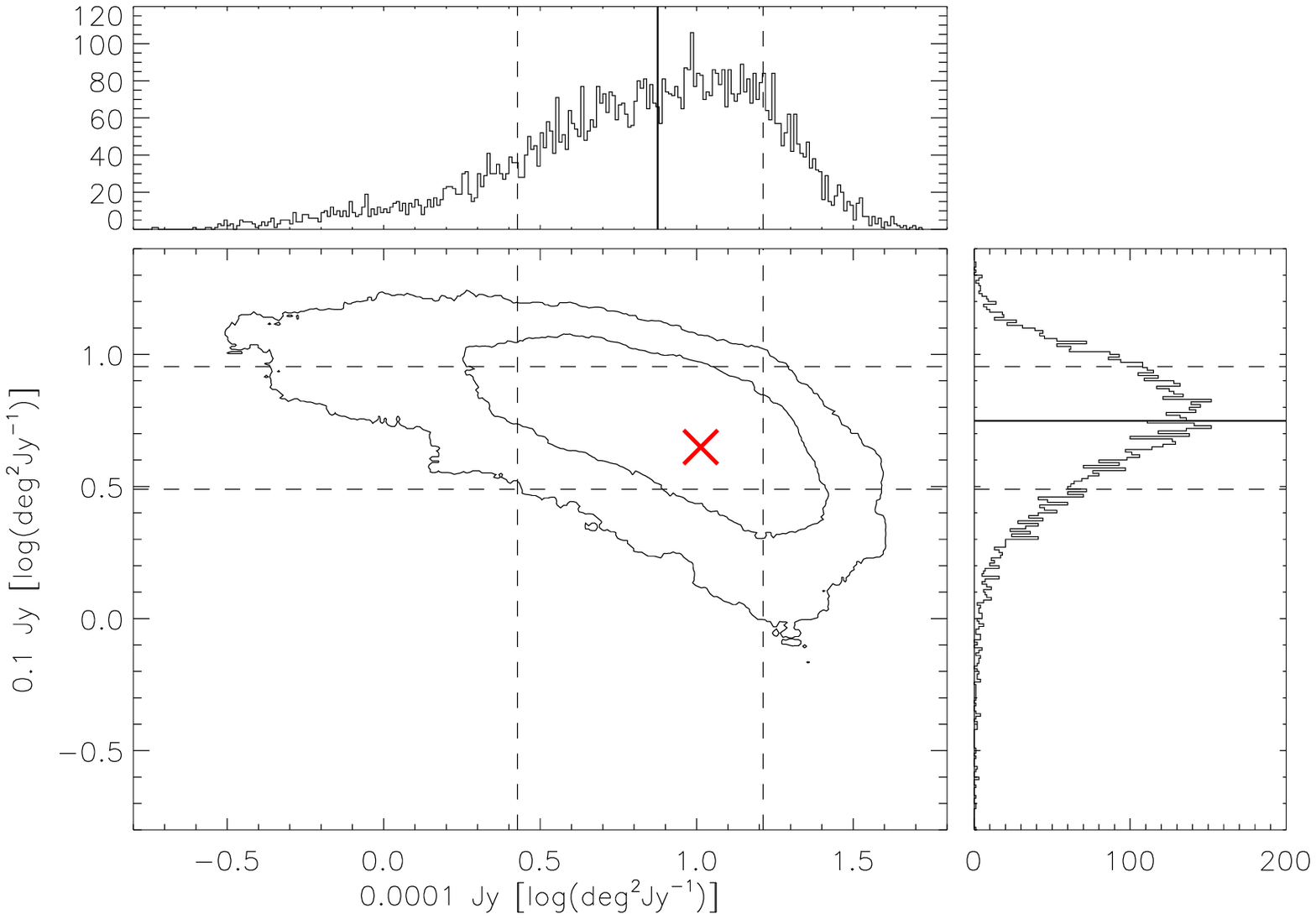}}
{\includegraphics[width=8.cm,keepaspectratio]{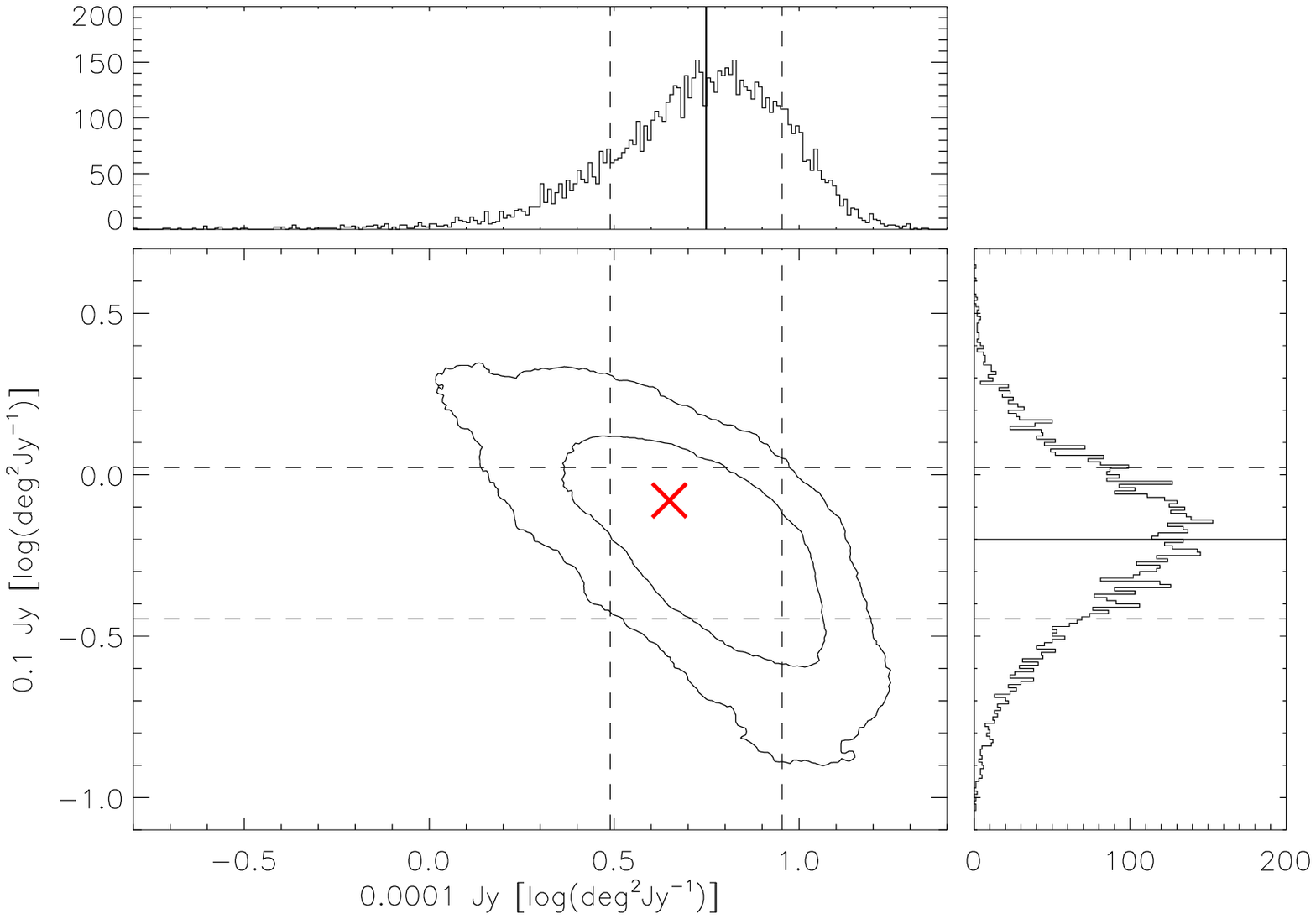}}\\
{\includegraphics[width=8.cm,keepaspectratio]{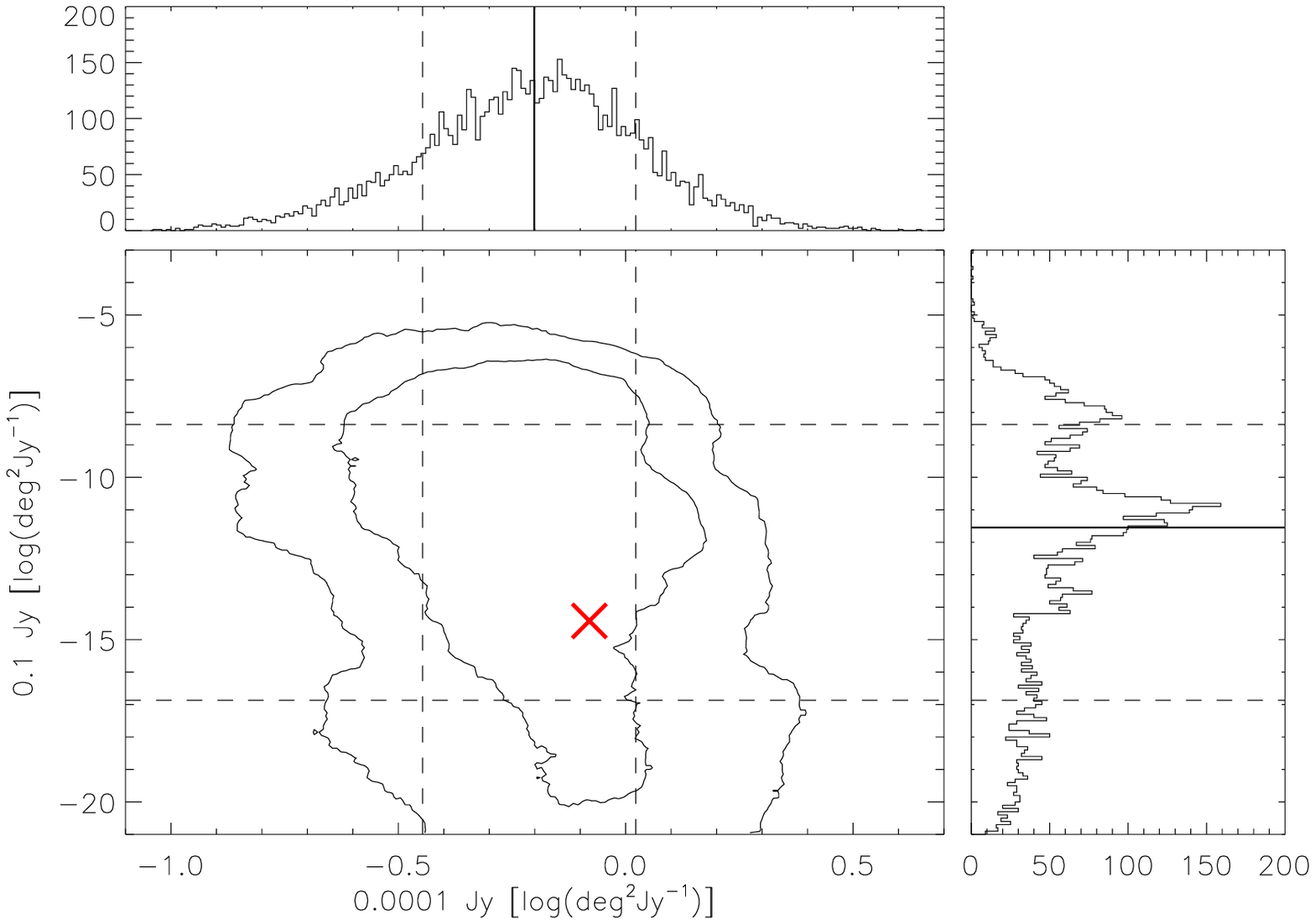}}
\caption{Likelihood distributions and contours for pairs of parameters associated with adjacent nodes at $250\,\micron$. The two curves in each panel represent $68\%$ and $95\%$ intervals and the solid and dashed lines represent the median and $1\sigma$ dispersion for each marginalized parameter: the distributions of the parameters are estimated from sampling the likelihood with MCMCMH. The red crosses mark the best-fit values.
}
\label{fig:paramsdistr}
\end{figure}

We have chosen to fit power-laws for differential number counts within predefined flux density bins, as described above. Five distinct power-laws are estimated (a total of six free parameters) for the differential number counts at $250\,\micron$ and four power-laws (five parameters) at both 350 and $500\,\micron$. The choice of flux densities for the boundary nodes is set by requiring them to be very far from the typical values constrained by BLAST, so that the result is independent of our particular choice. Best-fit number counts for the three wavelengths are presented in Fig.~\ref{fig:best_fit} and Table~\ref{tab:best_fit}. The uncertainties shown in the figure are computed from the $68\%$ confidence intervals on the marginalized distributions for each parameter separately. The marginal distributions (see Fig.~\ref{fig:paramsdistr}) have been estimated by sampling the likelihood around its maximum using a Markov Chain Monte Carlo Metropolis Hastings method (MCMCMH; \citealt{chib95}). The median values presented in Table~\ref{tab:best_fit} are not exactly the parameters of the best-fit model, due to non-Gaussian likelihoods around the maximum. Pearson correlation matrices for the parameters are given in Tables~\ref{tab:corrmat250}, \ref{tab:corrmat350} and  \ref{tab:corrmat500}. 

\begin{figure}

\centering
\includegraphics[width=13.cm,keepaspectratio]{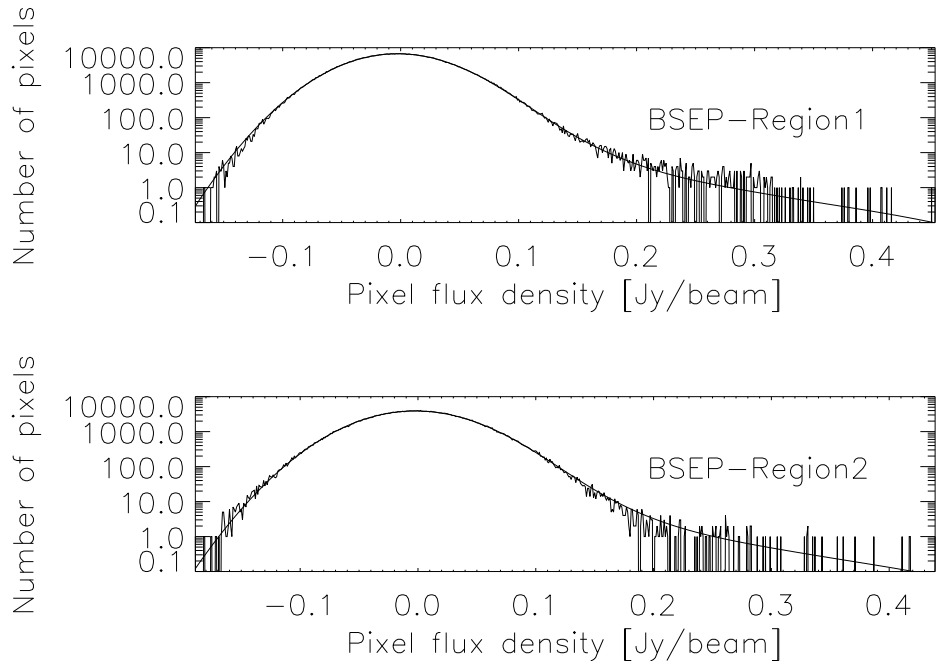}
\caption{Histograms of pixel values for two regions having different noise level, including about $\sim95\%$ of pixels of the BSEP map, compared with predictions of the best-fit model of the differential counts at $250\,\micron$. The corresponding plots at 350 and $500\,\micron$ look very similar. 
}
\label{fig:histograms}
\end{figure}

We compare the predicted histograms of the best-fit multi-power-law model with the actual histograms of the maps in Fig.~\ref{fig:histograms}. We plot the histograms of the two zones that include about $\sim95\%$ of the pixels and ignore the others here, since they only give weak additional constraints on the parameters. 

\cite{patanchon09} have demonstrated that, in a noise-dominated regime, it is useful to cross-correlate the maps with the beam kernel before {\itshape P(D)} analysis, even if this might not be the optimal choice. We have performed the analysis both convolving the maps with the same instrumental beam already used for the BGS maps and using a `matched filter', optimized for confusion-limited maps (see \S~\ref{sec:matched_filter}). As already pointed out in \S~\ref{sec:BLASTdata}, the BSEP maps are dominated by instrumental noise, thus the `matched filter' does not significantly improve the analysis: the differences between the two methods are much smaller that the uncertainties on the counts. 

The technique used in this paper also assumes that galaxies are randomly and independently distributed over the sky. However, significant correlations have been found in the background of the BGS maps \citep{viero09}, mainly due to clustering on scales larger than the BLAST beam. As already discussed in \cite{patanchon09}, the influence of clustering on the measurements of number counts is negligible with respect to the uncertainties. The clustering in BSEP is comparable to what is measured in BGS, despite having the outskirts of a local cluster (DRCG 0428$-$53 at $z=0.04$, \citealt{dressel80}) cover almost half of the map. As a test, we repeated the analysis on a sub-map that is free of cluster members. Our previous study on a galaxy cluster suggests that the contribution of the cluster to the submillimeter number counts may be significant in particular at the faint-end flux node in BSEP \citep{braglia10}, but the counts derived from the sub-map are consistent with those derived from the full map at the faint end of the distribution. At the bright end the counts from the sub-map are lower, but consistent within the small number statistics. In conclusion, the influence of the cluster can be considered negligible.

\begin{deluxetable}{c c c c c c c c c}
\tablecolumns{9}
\tablewidth{0pt}
\tablenum{1}
\tabletypesize{\footnotesize}
\tablecaption{Best-fit differential number counts}
\tablehead{
\multicolumn{3}{c}{$250\,\micron$} &  \multicolumn{3}{c}{$350\,\micron$} & \multicolumn{3}{c}{$500\,\micron$}  \\
\cline{1-3}
\cline{4-6}
\cline{7-9}
\colhead{Node} & \colhead{Best fit}   & \colhead{Marginal} &  \colhead{Node} & \colhead{Best fit}   & \colhead{Marginal} & \colhead{Node} & \colhead{Best fit}   & \colhead{Marginal} \\
\colhead{[Jy]} & \multicolumn{2}{c}{[$\log (\deg^2 \rm{Jy}^{-1}$)]}&\colhead{[Jy]} & \multicolumn{2}{c}{[$\log (\deg^2 \rm{Jy}^{-1}$)]}&\colhead{[Jy]} & \multicolumn{2}{c}{[$\log (\deg^2 \rm{Jy}^{-1}$)]}
}
\startdata   
$10^{-4}$ &   9.39  &   $<10.09$                  & $5 \times 10^{-5}$   &  11.64  &   $<12.10$               & $2.5 \times 10^{-5}$ &  12.43 &   $<12.81$\\
0.1       &   3.23  &    $3.27^{+0.11}_{-0.10}$   & 0.05                 &   3.86  &   $3.89^{+0.10}_{-0.11}$ & 0.03                 &  4.08  &   $4.24^{+0.13}_{-0.19}$\\
0.25      &   1.01  &    $0.88^{+0.34}_{-0.45}$   & 0.15                 &   1.16  &   $0.93^{+0.39}_{-0.55}$ & 0.08                 &  1.66  &   $1.24^{+0.66}_{-1.14}$\\
0.5       &   0.65  &    $0.75^{+0.21}_{-0.26}$   & 0.5                  &   0.31  &   $0.51^{+0.23}_{-0.28}$ & 0.25                 &  0.58  &   $0.67^{+0.32}_{-0.38}$\\
1.2       & $-0.080$&   $-0.20^{+0.22}_{-0.25}$   & 5                    & $-4.09$ &  $<-2.24$                & 2.5                  &$-11.17$&   $<-4.27$                           \\
10        &$-14.42$ &   $<-6.78$                  &                      &         &                          &                      &        &                                      \\
\enddata                                                    
\label{tab:best_fit}
\end{deluxetable}

\begin{deluxetable}{c r r r r r r}
\tablecolumns{7}
\tablewidth{0pt}
\tablenum{2}
\tabletypesize{\footnotesize}
\tablecaption{Pearson correlation matrix for the parameterized ${\rm d}N/{\rm d}S$ model at $250\,\micron$}
\tablehead{
\colhead{Node [Jy]} & \colhead{$10^{-4}$}   & \colhead{0.1} & \colhead{0.25} & \colhead{0.5} & \colhead{1.2} & \colhead{10}
}
\startdata   
$10^{-4}$ &  1.00  & $-0.94$ &  0.66  &$-0.29$  & $-0.002$ & $-0.02$  \\ 
0.1       &        &   1.00  &$-0.79$ &  0.38   & $-0.09$ &   0.05  \\   
0.25      &        &         &  1.00  &$-0.60$  &   0.21  & $-0.09$  \\  
0.5       &        &         &        &  1.00   & $-0.71$ &   0.16  \\   
1.2       &        &         &        &         &   1.00  & $-0.25$  \\  
10        &        &         &        &         &         &   1.00   \\  
\enddata 
\tablecomments{
Coefficients are computed as the covariance of the distributions of the parameters around the maximum of the likelihood. The distributions were derived using MCMCMH method. 
}                                                  
\label{tab:corrmat250}
\end{deluxetable}

\begin{deluxetable}{c r r r r r}
\tablecolumns{6}
\tablewidth{0pt}
\tablenum{3}
\tabletypesize{\footnotesize}
\tablecaption{Pearson correlation matrix for the parameterized ${\rm d}N/{\rm d}S$ model at $350\,\micron$}
\tablehead{
\colhead{Node [Jy]} & \colhead{$5 \times 10^{-5}$}   & \colhead{0.05} & \colhead{0.15} & \colhead{0.5} & \colhead{5}
}
\startdata   
$5 \times 10^{-5}$ &  1.00  &$-0.96$&  0.63 &$-0.43$&  0.07  \\         
0.05               &        &  1.00 &$-0.72$&  0.49 &$-0.09$  \\        
0.15               &        &       &  1.00 &$-0.71$&  0.16   \\        
0.5                &        &       &       &  1.00 &$-0.42$  \\        
5                  &        &       &       &       &  1.00    \\       
\enddata
\tablecomments{
Coefficients are computed as the covariance of the distributions of the parameters around the maximum of the likelihood. The distributions were derived using MCMCMH method.
}                                                  
\label{tab:corrmat350}
\end{deluxetable}

\begin{deluxetable}{c r r r r r}
\tablecolumns{6}
\tablewidth{0pt}
\tablenum{4}
\tabletypesize{\footnotesize}
\tablecaption{Pearson correlation matrix for the parameterized ${\rm d}N/{\rm d}S$ model at $500\,\micron$}
\tablehead{
\colhead{Node [Jy]} & \colhead{$2.5 \times 10^{-5}$}   & \colhead{0.03} & \colhead{0.08} & \colhead{0.25} & \colhead{2.5}
}
\startdata   
$2.5 \times 10^{-5}$ &  1.00  &$-0.96$&  0.67 &$-0.44$&  0.02   \\     
0.03                 &        &  1.00 &$-0.73$&  0.50 &$-0.01$   \\    
0.08                 &        &       &  1.00 &$-0.67$&  0.10   \\     
0.25                 &        &       &       &  1.00 &$-0.26$   \\    
2.5                  &        &       &       &       &  1.00    \\    
\enddata  
\tablecomments{
Coefficients are computed as the covariance of the distributions of the parameters around the maximum of the likelihood. The distributions were derived using MCMCMH method. 
}                                                 
\label{tab:corrmat500}
\end{deluxetable}

\subsection{Comparison with other measurements}
\label{sec:datacomp}
The number counts provided in this work can be compared with previously published counts in the same bands. All the existing differential number counts come from very recent maps provided by BLAST and {\itshape Herschel}/SPIRE. In Fig.~\ref{fig:best_fit} we compare our best-fit estimate for the differential number counts with the published counts by \cite{patanchon09}, \cite{oliver10} and \cite{clements10}. While the \cite{patanchon09} counts are derived from a statistical analysis of the map, \cite{oliver10} and \cite{clements10} extract candidate sources and apply corrections to the raw number counts for Eddington bias, reliability and incompleteness. Despite these differences, the number counts from these studies are in very good agreement with our results.

The sky coverage delivered by the BSEP map provides further constraints on the number counts at the bright end of the distribution. 

\subsection{Comparison with models}
\label{sec:modelscomp}
\begin{figure}

\centering
{\includegraphics[width=8.cm,keepaspectratio]{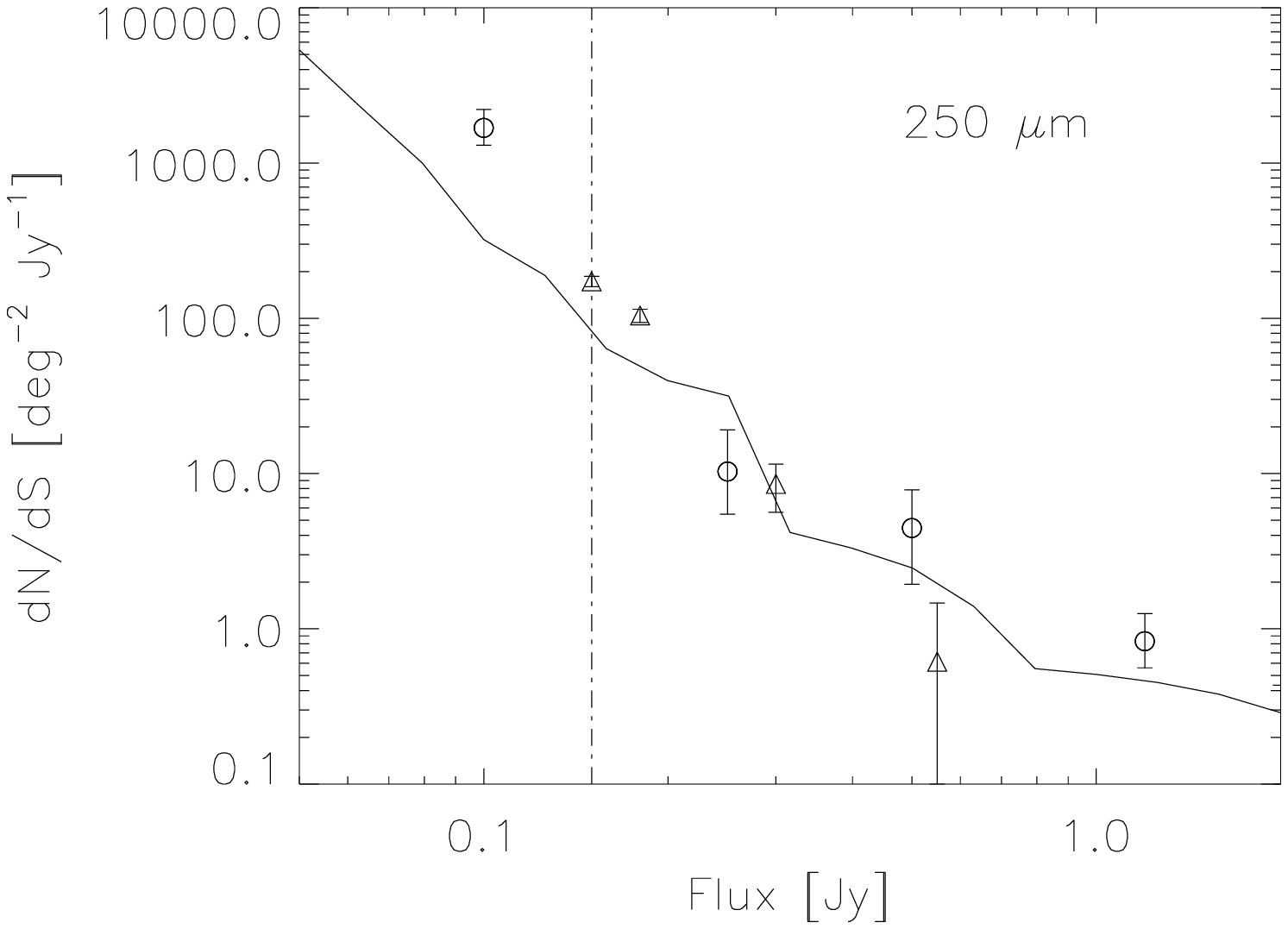}}
{\includegraphics[width=8.cm,keepaspectratio]{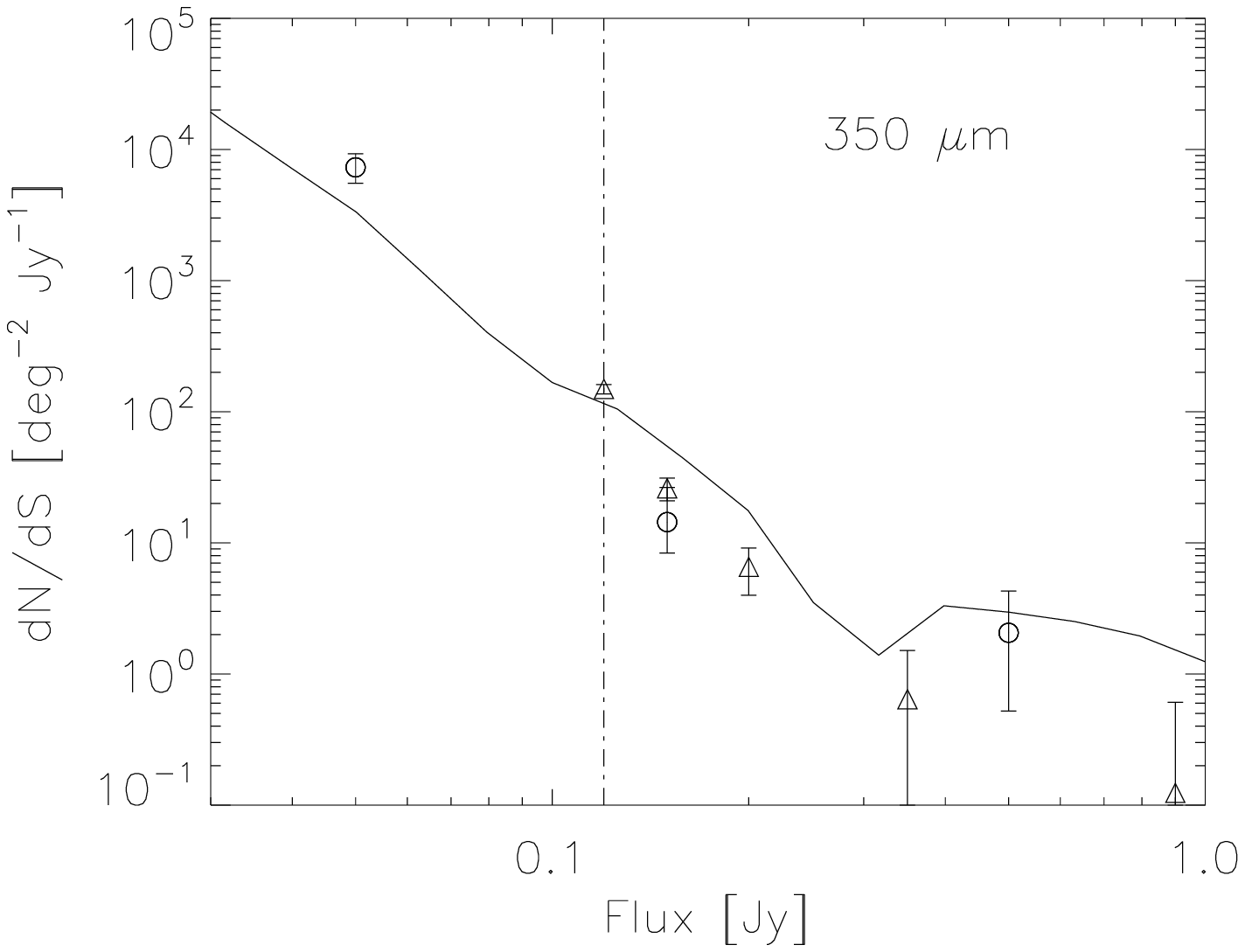}}\\
{\includegraphics[width=8.cm,keepaspectratio]{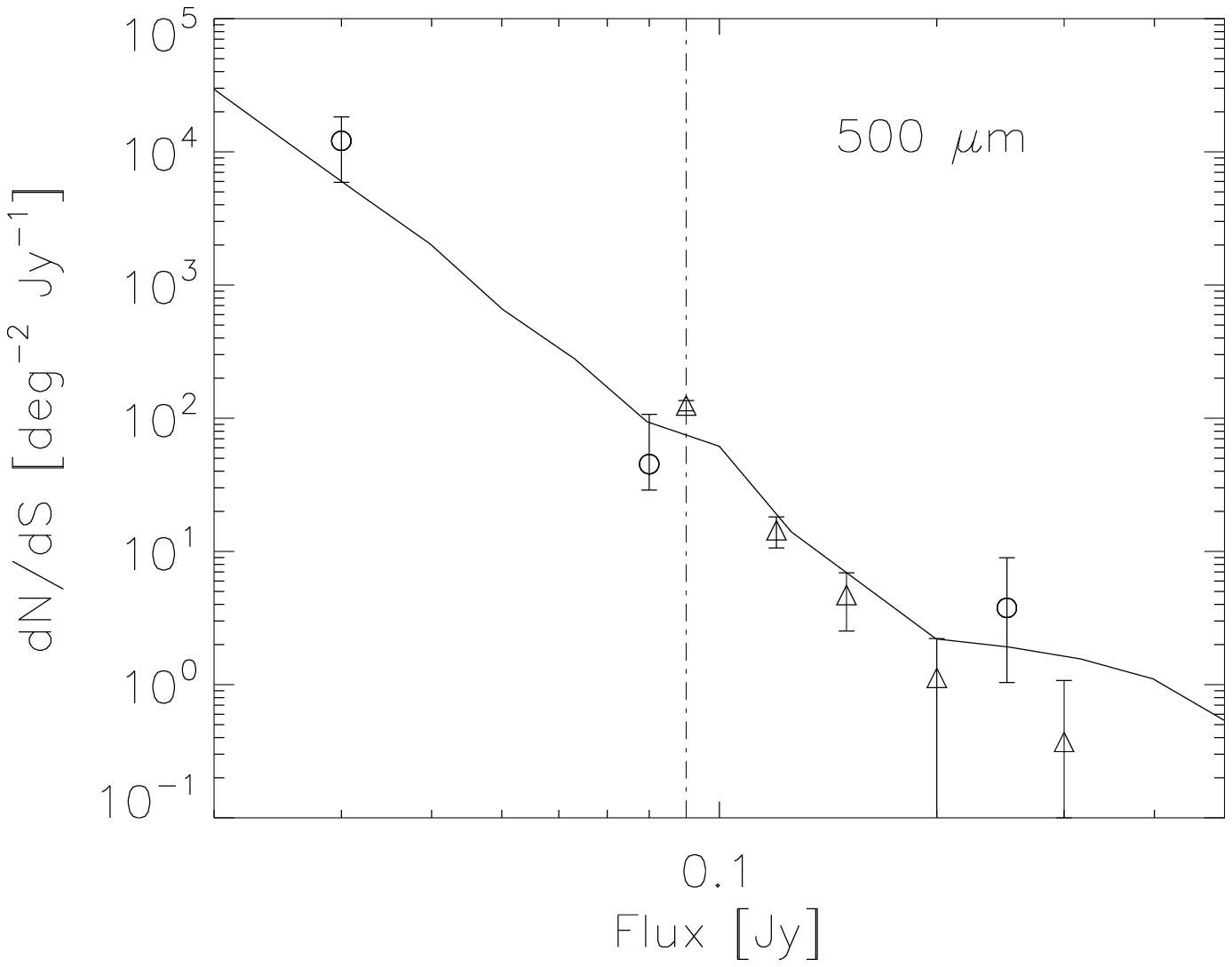}}
\caption{Best-fit differential number counts for the three BLAST wavelengths ({\itshape circles}) compared with a realization of the model of \cite{valiante09} ({\itshape solid line}). The plots also show the cruder estimate of the number counts derived directly from the $\geq 4\sigma$ catalogs after attempting to correct for completeness and false detection rate ({\itshape triangles}). The vertical lines define the flux limits of the catalogs. Note that the error bars are not marginalized in the same way as for the {\itshape P(D)} counts, so that it is not possible to make a simple comparison.
}
\label{fig:counts_mod}
\end{figure}

Fig.~\ref{fig:counts_mod} compares our best-fit counts with the model of \cite{valiante09}. As pointed out by \cite{oliver10}, this model has the distinct advantage, compared with most published models, of being able to fit the rise in the counts for fluxes in the range $10-100\,\rm{mJy}$. This could suggest that galaxies at high redshift have on average a colder temperature than local objects with the same luminosity. However, with suitable changes in the luminosity functions, it may be possible to make other models fit the observed counts, so it is premature to conclude that the spectral energy distributions need revision. Additionally, the model is able to reproduce the measured counts within the error bars in the range $50-1000\,\rm{mJy}$, as explored in these new BSEP data.  

\section{Source Catalogs}
\label{sec:cat}

We have compiled a catalog of point sources with flux $S\geq 3\sigma$ for each band (see Tables~\ref{tab:BLASTcat250}, \ref{tab:BLASTcat350}, \ref{tab:BLASTcat500}) using a source-finding algorithm which selects the peaks in a smoothed map produced by the noise-weighted convolution of the image with the telescope PSF \citep{devlin09}. The adopted $\sigma$ is the total noise of the map, $\sigma_{\rm t}$, defined in \S~\ref{sec:BLASTdata}. 
The source lists were synthesized into a common catalog using a procedure which accounts for the significance and positional uncertainty of the counterparts in each band. The radius of the $1\sigma$ positional error circle, $\sigma_{\rm p}$, for a submillimeter galaxy in a catalog with signal to noise $\mu$ which has not been corrected for the Eddington bias type of flux boosting can be expressed as
\begin{equation}
\sigma_{\rm p} = 0.9\theta[\mu^2-(2\alpha +4)]^{-1/2}
\end{equation}
for power-law counts of the form $N(>S)\propto S^{-\alpha}$, where $\theta$ is the full width at half maximum (FWHM) of the telescope beam \citep{ivison07}. Using this formula, error circles were calculated assuming $\alpha=2$, as obtained by fitting the number counts with a single power-law \citep{patanchon09}. A minimum $1\sigma$ error circle of $5^{\prime\prime}$ was imposed, equal to the $1\sigma$ pointing uncertainty of the maps.

The combined catalog is comprised of all sources with a significance $\geq 4\sigma$ in at least one band. Sources from other bands are considered to be matches if they are located within twice the radius of their respective error circles added in quadrature. Positions of sources in the resulting combined catalog were computed by averaging all the positions, weighted by $\sigma_{\rm p}^{-2}$.

The number of sources detected at $\geq 4\sigma$ by band is 132, 89 and 61 at 250, 350 and $500\,\micron$, respectively (see Table~\ref{tab:catalogs}). Table~\ref{tab:BLASTcat} lists the coordinates and 250, 350 and $500\,\micron$ flux densities and uncertainties for the 232 sources of the combined catalog.

\subsection{Completeness and false detection rates}

\begin{figure}

\centering
{\includegraphics[width=8.cm,keepaspectratio]{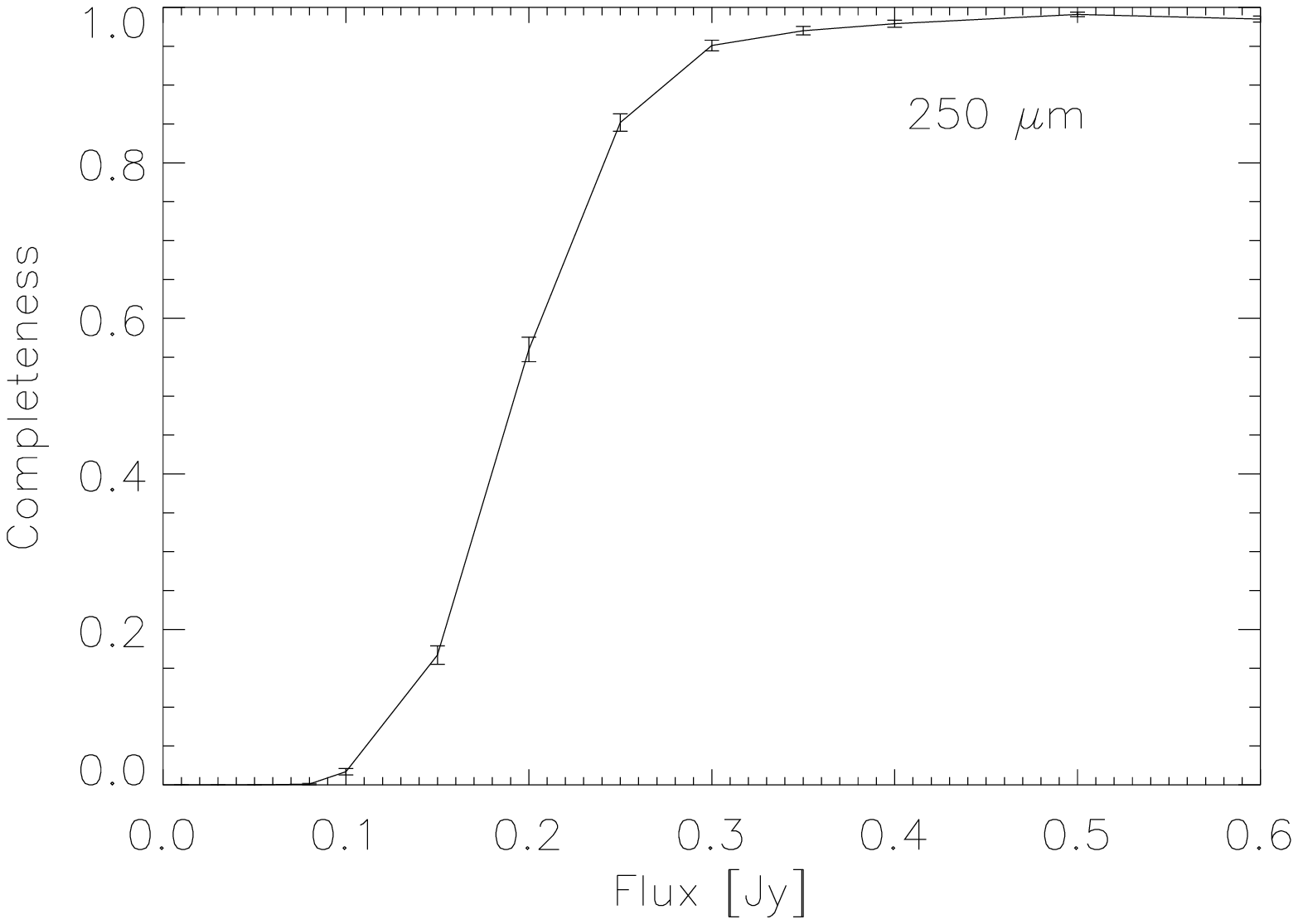}}
{\includegraphics[width=8.cm,keepaspectratio]{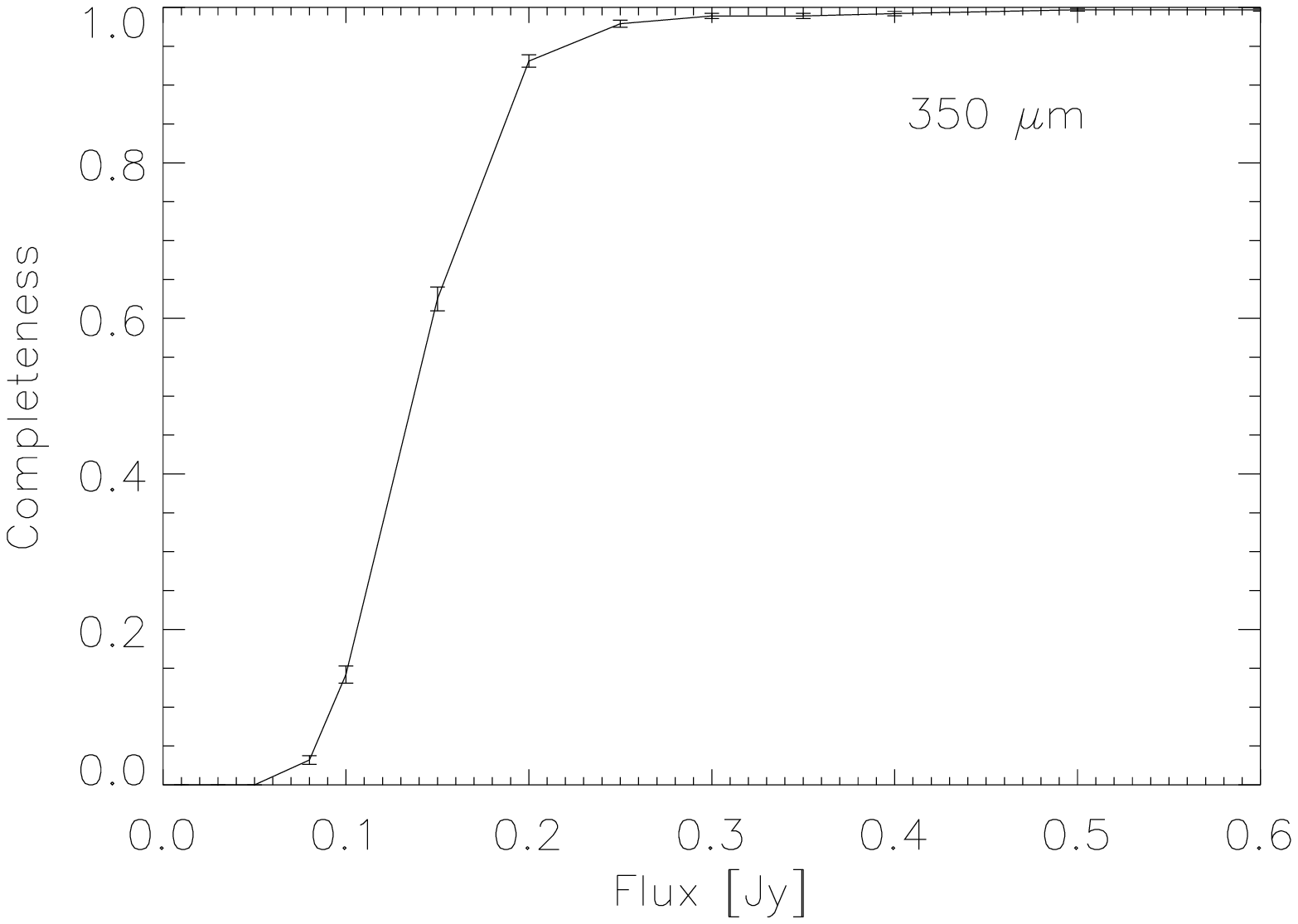}}\\
{\includegraphics[width=8.cm,keepaspectratio]{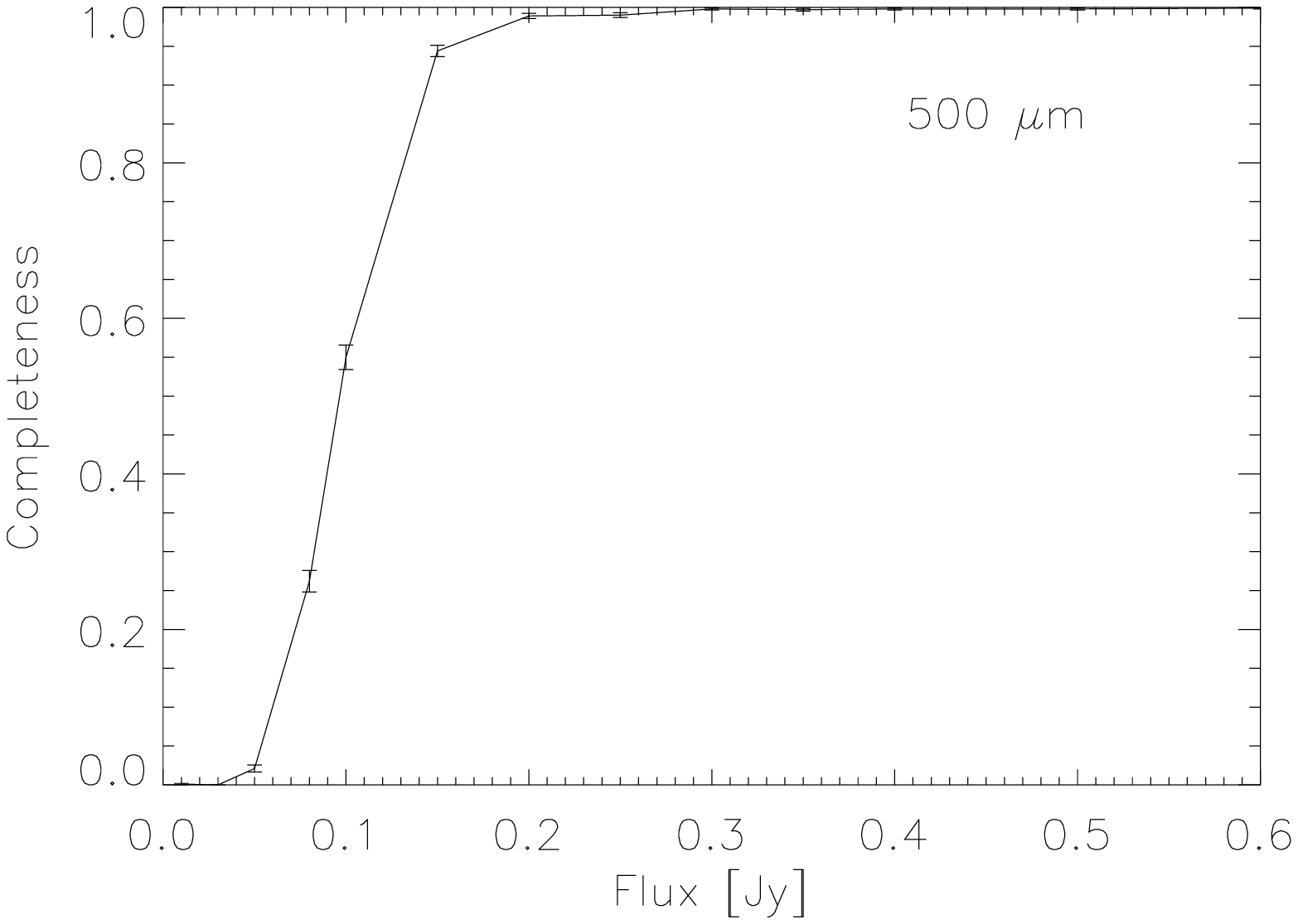}}
\caption{Catalog completeness at 250, 350 and $500\,\micron$. The error bars are estimated from $1\sigma$ binomial uncertainties.
}
\label{fig:completeness}
\end{figure}

`Completeness' is the probability of detecting a source of a given intrinsic flux density, given the depth of the survey and the source extraction algorithm. We estimate completeness by adding to the initial maps 1000 artificial point sources at random positions and calculating their recovery rates after performing the same source extraction procedure as for the real data. The artificial point sources are modeled as the PSF scaled by the flux density. To avoid blending the simulated sources together or with the existing sources, every simulated source is injected at a distance larger than 2.5 times the half width at half maximum from any real sources and from any other simulated ones. In principle this procedure introduces a bias which we estimate to be well under $1\%$. The number density of artificial sources is such that it does not appreciably change the noise properties of the maps. A source was considered to be detected if there was a detection within a circle centered on the source and within the radius of the FWHM of the corresponding wavelength. The catalog completeness as a function of intrinsic flux density is shown in Fig.~\ref{fig:completeness}. Table~\ref{tab:catalogs} gives the $50\%$, $80\%$ and $95\%$ completeness flux densities at each wavelength.

We calculate the false detection rates (FDRs) by performing the extraction algorithm on 100 simulated noise realization maps. This is a conservative overestimate of the FDRs and has the advantage of being model-independent \citep{perera08}. FDR values are reported in Table~\ref{tab:catalogs} for sources with a significance of $\geq 4\sigma$. At fluxes $\geq 5\sigma$ they are all consistent with zero.

Fig.~\ref{fig:counts_mod} shows the number counts derived from the $\geq 4\sigma$ catalogs after correcting for completeness and false detection rate. We can see that the catalog provides counts in good reasonably agreement with those derived by the {\itshape P(D)} analysis, but for a restricted flux density range.

\subsection{Matched filter}
\label{sec:matched_filter}

We have repeated the source extraction procedure along with completeness and false detection rate tests on maps which have been convolved with the `matched filter', calculated following the algorithm of \cite{chapin10} instead of the PSF. This filter is optimized to maximize the S/N of individual point sources in the presence of noise: it significantly reduces the confusion noise and slightly increases the white instrumental noise. This filter should perform significantly better as maps become more dominated by confusion.

Since the BSEP maps are dominated by instrumental noise, we find little benefit in the use of the `matched filter', as shown in Table~\ref{tab:catalogsMF}. Even though we identify a larger number of detected sources using the `matched filter', the fraction of false detections is also larger. For this reason, we prefer to publish the more conservative, and easily reproducible, source lists derived by filtering the maps with the PSF. 

\subsection{Extended sources}
\label{sec:extended}

The technique we have used to estimate the flux densities of the BLAST sources is not accurate in the case of extended sources. In the BSEP map there is one clearly extended source, identified with the galaxy NGC~1617 and detected as BLAST~J043137--543604 in our catalog. We have performed aperture photometry on this source and have estimated the uncertainty on the flux density taking into account both instrumental noise and calibration uncertainties. The measured values are reported in Table~\ref{tab:BLASTcat}.

\section{Conclusions}
We used BLAST to image a $9\,{\rm deg}^2$ field near the SEP at 250, 350 and $500\,\micron$, achieving median $1\sigma$ depths of 36.0, 26.4 and $18.4\,{\rm mJy}$ at each wavelength. We have identified 132, 89 and 61 sources with S/N $\geq 4\sigma$ at 250, 350 and $500\,\micron$, respectively. These have been compiled into a multi-wavelength catalog of 232 sources with a significance $\geq 4\sigma$ in at least one BLAST band. 

Using the {\itshape P(D)} technique, we have measured the differential number counts up to 1.2, 0.5 and $0.25\,{\rm Jy}$ at 250, 350 and $500\,\micron$, respectively. The new measurements agree with previous results from BLAST and more recent results from SPIRE and, thanks to the large area observed, give improved constrains at the bright end of the counts.

We have released the BLAST maps and catalogs to the public at http://blastexperiment.info.

\section{Acknowledgements}
BLAST acknowledges the support of NASA through grant numbers NAG5-12785, NAG5-13301, and NNGO-6GI11G, the NSF Office of Polar Programs, the Canadian Space Agency, the Natural Sciences and Engineering Research Council (NSERC) of Canada, and the UK Science and Technology Facilities Council (STFC). This research has been enabled by the use of WestGrid computing resources.

\begin{deluxetable}{c c c c c c c c}
\tablecolumns{8}
\tablewidth{0pt}
\tablenum{5}
\tabletypesize{\small}
\tablecaption{BLAST-SEP maps and catalogs}
\tablehead{
\colhead{Band} & \colhead{$\sigma_{\rm i}$} & \colhead{$\sigma_{\rm c}$} & \colhead{Number} & \colhead{False} & \colhead{$50\%$} & \colhead{$80\%$} & \colhead{$95\%$} \\
\colhead{} & \colhead{} & \colhead{} & \colhead{of sources} & \colhead{detections} & \colhead{completeness} & \colhead{completeness} & \colhead{completeness} \\
\colhead{[$\micron$]} & \colhead{[mJy]} & \colhead{[mJy]} & \colhead{$>4\sigma$} & \colhead{$>4\sigma$} & \colhead{[mJy]} & \colhead{[mJy]} & \colhead{[mJy]} 
}
\startdata   
250       & 36.0 & 20.6 & 132 & 2.28 & 192 & 241 & 299    \\
350       & 26.4 & 18.2 &  89 & 0.94 & 137 & 179 & 220    \\
500       & 18.4 & 15.2 &  61 & 0.14 &  97 & 132 & 157    \\
\enddata                                                  
\label{tab:catalogs}
\end{deluxetable}

\begin{deluxetable}{c c c c c c c c}
\tablecolumns{8}
\tablewidth{0pt}
\tablenum{6}
\tabletypesize{\small}
\tablecaption{BLAST-SEP maps and catalogs with Matched Filter}
\tablehead{
\colhead{Band} & \colhead{$\sigma_{\rm i}$} & \colhead{$\sigma_{\rm c}$} & \colhead{Number} & \colhead{False} & \colhead{$50\%$} & \colhead{$80\%$} & \colhead{$95\%$} \\
\colhead{} & \colhead{} & \colhead{} & \colhead{of sources} & \colhead{detections} & \colhead{completeness} & \colhead{completeness} & \colhead{completeness} \\
\colhead{[$\micron$]} & \colhead{[mJy]} & \colhead{[mJy]} & \colhead{$>4\sigma$} & \colhead{$>4\sigma$} & \colhead{[mJy]} & \colhead{[mJy]} & \colhead{[mJy]}
}
\startdata   
250       & 38.2 & 16.5 & 145 & 7.53 & 194 & 241 & 297   \\
350       & 28.9 & 14.2 & 100 & 6.80 & 144 & 184 & 226    \\
500       & 20.7 & 11.6 &  90 & 3.13 &  99 & 134 & 163    \\
\enddata                                                  
\label{tab:catalogsMF}
\end{deluxetable}

\clearpage

\section{Appendix A: data tables}
The following data tables are provided in the Appendix: the lists of submillimeter sources with flux $\geq 3\sigma$ at $250\,\micron$ (Table~\ref{tab:BLASTcat250}), $350\,\micron$ (Table~\ref{tab:BLASTcat350}) and $500\,\micron$ (Table~\ref{tab:BLASTcat500}); the multi-wavelength combined catalog of sources with a significance  $\geq 4\sigma$ in at least one band (Table~\ref{tab:BLASTcat}). Data tables are published in their entirety in the electronic edition of the {\itshape Astrophysical Journal Supplement Series}. The first 25 entries are shown here for guidance regarding their form and content.

\begin{deluxetable}{c c c c c c c}
\tablecolumns{7}
\tablewidth{0pt}
\tablenum{A1}
\tabletypesize{\footnotesize}
\tablecaption{Catalog of BLAST $250\,\micron$ sources in SEP with significance $\geq 3\sigma$. Flux densities are not corrected for Eddington-type bias.}
\tablehead{
\colhead{ID} & \colhead{BLAST ID}   & \colhead{RA} & \colhead{Dec} & \colhead{$S_{250}$} & \colhead{$\sigma_{250}$} & \colhead{S/N} \\
\colhead{}   & \colhead{}           & \colhead{(J2000)}& \colhead{(J2000)} & \colhead{Jy}        & \colhead{Jy}           & \colhead{}       \\
}
\startdata
    1 &  BLAST250 J043138--543604 &   67.909134 & --54.601269 &       $2.05^1$ &       $0.21^1$ &     $9.76$          \\   
    2 &  BLAST250 J043516--541902 &   68.818535 & --54.317352 &       0.818 &       0.041 &       19.77 \\
    3 &  BLAST250 J043126--542507 &   67.860474 & --54.418732 &       0.698 &       0.043 &       16.15 \\
    4 &  BLAST250 J043422--535358 &   68.591782 & --53.899590 &       0.549 &       0.041 &       13.40 \\
    5 &  BLAST250 J044936--535427 &   72.400566 & --53.907761 &       0.515 &       0.040 &       12.78 \\
    6 &  BLAST250 J042755--550332 &   66.980072 & --55.058910 &       0.716 &       0.053 &       13.61 \\
    7 &  BLAST250 J045412--532127 &   73.550888 & --53.357536 &       0.482 &       0.041 &       11.78 \\
    8 &  BLAST250 J045053--531233 &   72.724892 & --53.209232 &       0.482 &       0.042 &       11.56 \\
    9 &  BLAST250 J043424--544132 &   68.601288 & --54.692341 &       0.478 &       0.042 &       11.44 \\
   10 &  BLAST250 J044801--532609 &   72.005684 & --53.435989 &       0.467 &       0.043 &       10.97 \\
   11 &  BLAST250 J045045--533144 &   72.687759 & --53.528919 &       0.416 &       0.040 &       10.42 \\
   12 &  BLAST250 J045624--523057 &   74.101578 & --52.515938 &       0.361 &       0.040 &        9.10 \\
   13 &  BLAST250 J045313--524746 &   73.306427 & --52.796268 &       0.345 &       0.040 &        8.69 \\
   14 &  BLAST250 J044839--535429 &   72.165001 & --53.908199 &       0.339 &       0.041 &        8.36 \\
   15 &  BLAST250 J045307--525416 &   73.282074 & --52.904690 &       0.337 &       0.040 &        8.34 \\
   16 &  BLAST250 J042705--541554 &   66.773399 & --54.265018 &       0.333 &       0.040 &        8.23 \\
   17 &  BLAST250 J044706--524028 &   71.777534 & --52.674500 &       0.320 &       0.039 &        8.10 \\
   18 &  BLAST250 J045759--524129 &   74.499512 & --52.691544 &       0.331 &       0.041 &        8.18 \\
   19 &  BLAST250 J043553--542417 &   68.973793 & --54.404816 &       0.351 &       0.042 &        8.28 \\
   20 &  BLAST250 J043513--541230 &   68.807793 & --54.208553 &       0.329 &       0.041 &        8.04 \\
   21 &  BLAST250 J045443--530540 &   73.680298 & --53.094635 &       0.314 &       0.041 &        7.63 \\
   22 &  BLAST250 J045139--534459 &   72.916176 & --53.749763 &       0.305 &       0.040 &        7.55 \\
   23 &  BLAST250 J044853--523039 &   72.221237 & --52.511059 &       0.306 &       0.041 &        7.53 \\
   24 &  BLAST250 J044125--532834 &   70.357887 & --53.476337 &       0.318 &       0.042 &        7.59 \\
   25 &  BLAST250 J043346--534651 &   68.441719 & --53.781109 &       0.295 &       0.040 &        7.38 \\
\enddata
\vspace{0.5cm}\\
\begin{tabular}{ll}
$^1$ & These flux densities come from aperture photometry. See \S~\ref{sec:extended} for details.
\end{tabular}                                                  
\label{tab:BLASTcat250}                                       
\end{deluxetable}

\begin{deluxetable}{c c c c c c c}
\tablecolumns{7}
\tablewidth{0pt}
\tablenum{A2}
\tabletypesize{\footnotesize}
\tablecaption{Catalog of BLAST $350\,\micron$ sources in SEP with significance $\geq 3\sigma$. Flux densities are not corrected for Eddington-type bias.}
\tablehead{
\colhead{ID} & \colhead{BLAST ID}   & \colhead{RA} & \colhead{Dec} & \colhead{$S_{350}$} & \colhead{$\sigma_{350}$} & \colhead{S/N} \\
\colhead{}   & \colhead{}           & \colhead{(J2000)}& \colhead{(J2000)} & \colhead{Jy}        & \colhead{Jy}             & \colhead{}    \\
}
\startdata
    1 &  BLAST350 J043137--543603 &   67.904358 & --54.601097 &       $1.09^1$ &       $0.12^1$ &       $9.08$  \\   
    2 &  BLAST350 J043516--541902 &   68.818398 & --54.317253 &       0.351 &       0.032 &       10.92 \\
    3 &  BLAST350 J042754--550331 &   66.975388 & --55.058666 &       0.472 &       0.044 &       10.80 \\
    4 &  BLAST350 J043125--542516 &   67.855537 & --54.421322 &       0.325 &       0.033 &        9.88 \\
    5 &  BLAST350 J043421--535408 &   68.591515 & --53.902260 &       0.280 &       0.031 &        8.93 \\
    6 &  BLAST350 J045412--532127 &   73.550613 & --53.357510 &       0.272 &       0.031 &        8.66 \\
    7 &  BLAST350 J044935--535437 &   72.396095 & --53.910458 &       0.248 &       0.032 &        7.80 \\
    8 &  BLAST350 J045044--533143 &   72.687477 & --53.528831 &       0.229 &       0.031 &        7.37 \\
    9 &  BLAST350 J044706--524018 &   71.777740 & --52.671692 &       0.204 &       0.031 &        6.60 \\
   10 &  BLAST350 J043424--544132 &   68.601440 & --54.692429 &       0.198 &       0.032 &        6.11 \\
   11 &  BLAST350 J045317--533816 &   73.321846 & --53.637840 &       0.185 &       0.031 &        5.87 \\
   12 &  BLAST350 J045053--531233 &   72.724693 & --53.209290 &       0.197 &       0.033 &        5.97 \\
   13 &  BLAST350 J045025--524125 &   72.607147 & --52.690285 &       0.180 &       0.031 &        5.77 \\
   14 &  BLAST350 J043624--542508 &   69.101585 & --54.419037 &       0.178 &       0.033 &        5.32 \\
   15 &  BLAST350 J043935--530926 &   69.898155 & --53.157463 &       0.217 &       0.039 &        5.53 \\
   16 &  BLAST350 J044839--535428 &   72.164772 & --53.908028 &       0.152 &       0.030 &        5.00 \\
   17 &  BLAST350 J044436--540946 &   71.150352 & --54.162781 &       0.165 &       0.032 &        5.10 \\
   18 &  BLAST350 J044853--523039 &   72.221313 & --52.510956 &       0.157 &       0.031 &        5.01 \\
   19 &  BLAST350 J044802--532608 &   72.010269 & --53.435829 &       0.172 &       0.033 &        5.14 \\
   20 &  BLAST350 J044040--542055 &   70.169220 & --54.348644 &       0.161 &       0.032 &        4.96 \\
   21 &  BLAST350 J045623--523106 &   74.097374 & --52.518547 &       0.148 &       0.031 &        4.80 \\
   22 &  BLAST350 J044944--525427 &   72.434944 & --52.907745 &       0.146 &       0.031 &        4.77 \\
   23 &  BLAST350 J044831--540229 &   72.132347 & --54.041489 &       0.154 &       0.032 &        4.85 \\
   24 &  BLAST350 J043512--541210 &   68.803497 & --54.202976 &       0.153 &       0.032 &        4.81 \\
   25 &  BLAST350 J043830--541840 &   69.628357 & --54.311268 &       0.153 &       0.032 &        4.77 \\
\enddata
\vspace{0.5cm}\\
\begin{tabular}{ll}
$^1$ & These flux densities come from aperture photometry. See \S~\ref{sec:extended} for details.
\end{tabular}                                                  
\label{tab:BLASTcat350}                                       
\end{deluxetable}

\begin{deluxetable}{c c c c c c c}
\tablecolumns{10}
\tablewidth{0pt}
\tablenum{A3}
\tabletypesize{\footnotesize}
\tablecaption{Catalog of BLAST $500\,\micron$ sources in SEP with significance $\geq 3\sigma$. Flux densities are not corrected for Eddington-type bias.}
\tablehead{
\colhead{ID} & \colhead{BLAST ID}   & \colhead{RA} & \colhead{Dec} & \colhead{$S_{500}$} & \colhead{$\sigma_{500}$} & \colhead{S/N} \\
\colhead{}   & \colhead{}           & \colhead{(J2000)}& \colhead{(J2000)} & \colhead{Jy}        & \colhead{Jy}           & \colhead{}     \\
}
\startdata
    1 &  BLAST500 J043137--543603 &   67.904343 & --54.601093 &       $0.395^1$ &       $0.055^1$ &       $7.18$ \\   
    2 &  BLAST500 J042756--550322 &   66.985207 & --55.056328 &       0.257 &       0.030 &        8.52 \\
    3 &  BLAST500 J043125--542506 &   67.855965 & --54.418583 &       0.167 &       0.025 &        6.73 \\
    4 &  BLAST500 J045056--531633 &   72.735146 & --53.275887 &       0.150 &       0.023 &        6.40 \\
    5 &  BLAST500 J043516--541912 &   68.818130 & --54.320080 &       0.147 &       0.024 &        6.21 \\
    6 &  BLAST500 J043521--550056 &   68.840088 & --55.015621 &       0.147 &       0.026 &        5.73 \\
    7 &  BLAST500 J044742--533019 &   71.926003 & --53.505398 &       0.130 &       0.024 &        5.36 \\
    8 &  BLAST500 J045600--524902 &   74.001236 & --52.817482 &       0.115 &       0.023 &        5.04 \\
    9 &  BLAST500 J042640--541057 &   66.669762 & --54.182514 &       0.150 &       0.027 &        5.56 \\
   10 &  BLAST500 J045046--533133 &   72.692162 & --53.526104 &       0.114 &       0.023 &        5.00 \\
   11 &  BLAST500 J045307--525357 &   73.281929 & --52.899330 &       0.118 &       0.023 &        5.06 \\
   12 &  BLAST500 J050244--525150 &   75.686989 & --52.864151 &       0.154 &       0.028 &        5.57 \\
   13 &  BLAST500 J045920--521402 &   74.836861 & --52.234043 &       0.116 &       0.023 &        4.96 \\
   14 &  BLAST500 J045107--525202 &   72.779480 & --52.867264 &       0.109 &       0.023 &        4.78 \\
   15 &  BLAST500 J044156--531041 &   70.485947 & --53.178085 &       0.115 &       0.024 &        4.82 \\
   16 &  BLAST500 J050215--523923 &   75.564407 & --52.656528 &       0.130 &       0.026 &        5.04 \\
   17 &  BLAST500 J043750--532126 &   69.459808 & --53.357227 &       0.120 &       0.025 &        4.89 \\
   18 &  BLAST500 J044350--525220 &   70.961983 & --52.872311 &       0.118 &       0.024 &        4.85 \\
   19 &  BLAST500 J043412--545017 &   68.550507 & --54.838169 &       0.121 &       0.025 &        4.89 \\
   20 &  BLAST500 J044310--542034 &   70.793472 & --54.343006 &       0.118 &       0.024 &        4.84 \\
   21 &  BLAST500 J045840--523546 &   74.668068 & --52.596275 &       0.120 &       0.025 &        4.83 \\
   22 &  BLAST500 J045442--530530 &   73.675896 & --53.091942 &       0.110 &       0.024 &        4.63 \\
   23 &  BLAST500 J044154--540351 &   70.475601 & --54.064274 &       0.104 &       0.023 &        4.49 \\
   24 &  BLAST500 J044232--535158 &   70.633949 & --53.866184 &       0.106 &       0.024 &        4.50 \\
   25 &  BLAST500 J043832--541811 &   69.633400 & --54.303066 &       0.107 &       0.024 &        4.48 \\
\enddata
\vspace{0.5cm}\\
\begin{tabular}{ll}
$^1$ & These flux densities come from aperture photometry. See \S~\ref{sec:extended} for details.
\end{tabular}                                                  
\label{tab:BLASTcat500}                                       
\end{deluxetable}

\begin{deluxetable}{c c c c c c c c c c}
\tablecolumns{10}
\tablewidth{0pt}
\tablenum{A4}
\tabletypesize{\footnotesize}
\tablecaption{Multi-wavelength catalog of BLAST sources in SEP with $\geq 4\sigma$ detection in at least one band. Flux densities are not corrected for Eddington-type bias. The non-detections are expressed as $3\sigma$ upper limits.}
\tablehead{
\colhead{ID} & \colhead{BLAST ID}   & \colhead{RA} & \colhead{Dec} & \colhead{$S_{250}$} & \colhead{$\sigma_{250}$} & \colhead{$S_{350}$} & \colhead{$\sigma_{350}$} & \colhead{$S_{500}$} & \colhead{$\sigma_{500}$}\\
\colhead{}   & \colhead{}           & \colhead{(J2000)}& \colhead{(J2000)} & \colhead{Jy}        & \colhead{Jy}           & \colhead{Jy}        & \colhead{Jy}           & \colhead{Jy}        & \colhead{Jy}         \\
}
\startdata
  1 &     BLAST J043137--543604 &   67.905945 &  --54.601154 &       $2.05^1$ &       $0.21^1$ &       $1.09^1$ &      $0.12^1$ &       $0.395^1$ &       $0.055^1$ \\   
  2 &     BLAST J043516--541904 &   68.818398 & --54.317852 &       0.818 &       0.041 &       0.351 &       0.032 &       0.147 &       0.024 \\
  3 &     BLAST J043125--542510 &   67.857567 & --54.419716 &       0.698 &       0.043 &       0.325 &       0.033 &       0.167 &       0.025 \\
  4 &     BLAST J043422--535403 &   68.591995 & --53.901043 &       0.549 &       0.041 &       0.280 &       0.031 &       0.094 &       0.023 \\
  5 &     BLAST J044935--535432 &   72.398254 & --53.909065 &       0.515 &       0.040 &       0.248 &       0.032 &       0.070 &       0.023 \\
  6 &     BLAST J042755--550329 &   66.979797 & --55.058105 &       0.716 &       0.053 &       0.472 &       0.044 &       0.257 &       0.030 \\
  7 &     BLAST J045412--532127 &   73.550751 & --53.357513 &       0.482 &       0.041 &       0.272 &       0.031 &       0.092 &       0.024 \\
  8 &     BLAST J045053--531232 &   72.724510 & --53.208916 &       0.482 &       0.042 &       0.197 &       0.033 &       0.085 &       0.024 \\
  9 &     BLAST J043424--544132 &   68.601357 & --54.692379 &       0.478 &       0.042 &       0.198 &       0.032 &    $<$0.074 &         $-$ \\
 10 &     BLAST J044801--532609 &   72.007538 & --53.435932 &       0.467 &       0.043 &       0.172 &       0.033 &       0.097 &       0.024 \\
 11 &     BLAST J045045--533142 &   72.688240 & --53.528496 &       0.416 &       0.040 &       0.229 &       0.031 &       0.114 &       0.023 \\
 12 &     BLAST J045624--523100 &   74.100220 & --52.516781 &       0.361 &       0.040 &       0.148 &       0.031 &    $<$0.071 &         $-$ \\
 13 &     BLAST J045313--524746 &   73.306450 & --52.796272 &       0.345 &       0.040 &       0.129 &       0.031 &    $<$0.069 &         $-$ \\
 14 &     BLAST J044839--535429 &   72.164917 & --53.908138 &       0.339 &       0.041 &       0.152 &       0.030 &    $<$0.069 &         $-$ \\
 15 &     BLAST J045307--525411 &   73.282059 & --52.903137 &       0.337 &       0.040 &       0.137 &       0.031 &       0.118 &       0.023 \\
 16 &     BLAST J042705--541554 &   66.773399 & --54.265018 &       0.333 &       0.040 &    $<$0.095 &         $-$ &    $<$0.072 &         $-$ \\
 17 &     BLAST J044706--524023 &   71.777649 & --52.673302 &       0.320 &       0.039 &       0.204 &       0.031 &       0.073 &       0.022 \\
 18 &     BLAST J045759--524129 &   74.499466 & --52.691570 &       0.331 &       0.041 &       0.131 &       0.032 &    $<$0.069 &         $-$ \\
 19 &     BLAST J043553--542417 &   68.973129 & --54.404781 &       0.351 &       0.042 &       0.117 &       0.033 &    $<$0.072 &         $-$ \\
 20 &     BLAST J043513--541228 &   68.807434 & --54.207802 &       0.329 &       0.041 &    $<$0.095 &         $-$ &       0.081 &       0.024 \\
 21 &     BLAST J045443--530537 &   73.679634 & --53.093685 &       0.314 &       0.041 &       0.132 &       0.032 &       0.110 &       0.024 \\
 22 &     BLAST J045139--534501 &   72.916527 & --53.750340 &       0.305 &       0.040 &    $<$0.094 &         $-$ &       0.073 &       0.023 \\
 23 &     BLAST J044853--523039 &   72.221260 & --52.511024 &       0.306 &       0.041 &       0.157 &       0.031 &    $<$0.071 &         $-$ \\
 24 &     BLAST J044126--532832 &   70.359184 & --53.475574 &       0.318 &       0.042 &       0.146 &       0.032 &    $<$0.073 &         $-$ \\
 25 &     BLAST J043346--534651 &   68.441719 & --53.781109 &       0.295 &       0.040 &    $<$0.096 &         $-$ &    $<$0.073 &         $-$ \\
\enddata
\vspace{0.5cm}\\
\begin{tabular}{ll}
$^1$ & These flux densities come from aperture photometry. See \S~\ref{sec:extended} for details.
\end{tabular}                                                  
\label{tab:BLASTcat}                                       
\end{deluxetable}


\clearpage

\begin{thebibliography}{999}
\bibitem[Bethermin et al.(2010)]{bethermin10} Bethermin, M., Dole, H., Cousin, M., Bavouzet, N. 2010, \aap, 516, 43
\bibitem[Braglia et al.(2010)]{braglia10} Braglia, F.G., et al. 2010, \mnras, submitted (arXiv:1003.2629)
\bibitem[Chapin et al.(2010)]{chapin10} Chapin, E.L., et al. 2010, \mnras, submitted (arXiv:1003.2647)
\bibitem[Chib \& Greenberg (1995)]{chib95} Chib, S., Greenberg, E. 1995, Am. Stat., 49, 327
\bibitem[Clements et al.(2010)]{clements10} Clements, D., et al. 2010, \aap, 518, 8L
\bibitem[Devlin et al.(2009)]{devlin09} Devlin, M.J., et al. 2009, \nat, 458, 737
\bibitem[Dickinson et al.(2007)]{dickinson07} Dickinson, M., et al. 2007, AAS, 211, 5216
\bibitem[Dressel (1980)]{dressel80} Dressel, A. 1980, \apjs, 42, 565
\bibitem[Fixsen et al.(1998)]{fixsen98} Fixsen, D.J., Dwek, E., Mather, J.C., Bennett, C.L., Shafer, R.A. 1998, \apj, 508, 123
\bibitem[Hauser \& Dwek (2001)]{hauser01} Hauser, M.G., Dwek, E, 2001, \araa, 39, 249
\bibitem[Ivison et al.(2007)]{ivison07} Ivison, R.J., et al. 2007, \mnras, 380, 199
\bibitem[Malek et al.(2010)]{malek10} Malek, K., Pollo, A., Takeuchi, T.T., Bienias, P., Shirahata, M., Matsuura, S., Kawada, M. 2009, \aap, 514, 11
\bibitem[Marsden et al.(2009)]{marsden09} Marsden, G., et al. 2009, \apj, 707, 1729
\bibitem[Matsuhara et al.(2006)]{matsuhara06} Matsuhara, H., et al. 2006, \pasj, 58, 673
\bibitem[Matsuura et al.(2010)]{matsuura10} Matsuura, S., et al. 2010, \apj, submitted (arXiv:1002.3674)
\bibitem[Oliver et al.(2010)]{oliver10} Oliver, S.J., et al. 2010, \aap, 518, 21L
\bibitem[Pascale et al.(2008)]{pascale08} Pascale, E., et al. 2008, \apj, 681, 400
\bibitem[Pascale et al.(2009)]{pascale09} Pascale, E., et al. 2009, \apj, 707, 1740
\bibitem[Patanchon et al.(2008)]{patanchon08} Patanchon, G., et al. 2008, \apj, 681, 708
\bibitem[Patanchon et al.(2009)]{patanchon09} Patanchon, G., et al. 2009, \apj, 707, 1750
\bibitem[Perera et al.(2008)]{perera08} Perera, T.A, et al. 2008, \mnras, 391, 1227
\bibitem[Schlegel et al.(1998)]{schlegel98} Schlegel, D.J., Finkbeiner, D.P., Davis, M. 1998, \apj, 500, 525
\bibitem[Scott et al.(2010)]{scott10} Scott, K., et al. 2010, \apjs, accepted, (arXiv:1007.0038)
\bibitem[Sutherland \& Saunders(1992)]{sutherland92} Sutherland, W., Saunders, W. 1992, \mnras, 259, 413
\bibitem[Truch et al.(2009)]{truch09} Truch, M.D.P., et al. 2009, \apj, 707, 1723
\bibitem[Valiante et al.(2009)]{valiante09} Valiante, E., Lutz, D., Sturm, E., Genzel, R., Chapin, E.L. 2009, \apj, 701, 1814
\bibitem[Viero et al.(2009)]{viero09} Viero, M., et al. 2009, \apj, 707, 1766
\end{thebibliography}
\end{document}